\documentclass[aps, prl,reprint,a4paper,onecolumn]{revtex4-1}

\usepackage[utf8]{inputenc}
\usepackage{graphicx}
\usepackage{amsmath}
\usepackage{amssymb}
\usepackage{color}
\usepackage{soul}

\begin{document}

\title{A phase field approach to trabecular bone remodeling}

\author{S. Aland}
\affiliation{Faculty of Informatics/Mathematics, HTW Dresden, Germany}
\author{F. Stenger}
\affiliation{Institut f\"ur Wissenschaftliches Rechnen, TU Dresden, Germany}
\author{R. M\"uller}
\affiliation{Center for Information Services and High Performance Computing, TU Dresden, Dresden, Germany}
\author{M. Kampschulte}
\affiliation{Diagnostische Radiologie, Justus-Liebig-Universit\"at Gie{\ss}en, Germany}
\author{A.C. Langheinrich}
\affiliation{Berufsgenossenschaftliche Unfallklinik Frankfurt am Main, Germany}
\author{T. El Khassawna}
\affiliation{Experimental Trauma Surgery, Justus-Liebig University of Giessen, Germany}
\author{C. Heiss}
\affiliation{Department of Trauma, Hand and Reconstructive Surgery, University Hospital of Giessen-Marburg GmbH, Campus: Giessen, Germany}
\author{A. Deutsch}
\affiliation{Center for Information Services and High Performance Computing, TU Dresden, Dresden, Germany}
\author{A. Voigt}  
\affiliation{Institut f\"ur Wissenschaftliches Rechnen, TU Dresden, Germany}
\affiliation{Dresden Center for Computational Materials Science (DCMS), Dresden, Germany}
\affiliation{Center for Systems Biology Dresden (CSBD), Dresden, Germany}

\begin{abstract}
We introduce a continuous modeling approach which combines elastic responds of the trabecular bone structure, the concentration of signaling mole-cules within the bone and a mechanism how this concentration at the bone surface is used for local bone formation and resorption. In an abstract setting bone can be considered as a shape changing structure. For similar problems in materials science phase field approximations have been established as an efficient computational tool. We adapt such an approach for trabecular bone remodeling. It allows for a smooth representation of the trabecular bone structure and drastically reduces computational costs if compared with traditional micro finite element approaches. We demonstrate the advantage of the approach within a minimal model. We quantitatively compare the results with established micro finite element approaches on simple geometries and consider the bone morphology within a bone segment obtained from $\mu$CT data of a sheep vertebra with realistic parameters. 
\end{abstract}

\maketitle

\section{Introduction}

\label{sec1}

\begin{figure}[h]
\centering
\includegraphics[width=.9\linewidth]{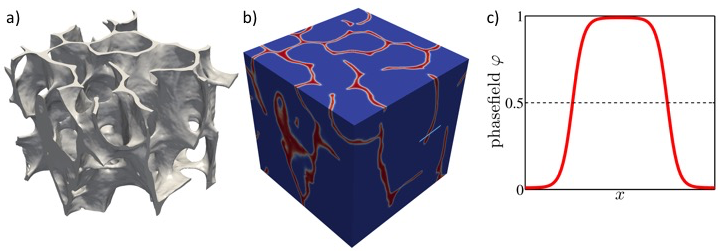}
\caption{(a) Trabecular structure from sheep tomography data represented by level surface of the phase field function. (b) phase field function in a box domain. (c) phase field function along the light blue line in (b) with $\phi = 1$ representing the bone structure.}
\label{fig1}
\end{figure}

Bone undergoes a continuous renewal process, which helps to maintain its mechanical performance and allows for adaptation to changes in mechanical requirements. Different cells are involved in this remodeling process: osteoclasts, which resorb bone, and osteoblasts, which deposit bone. The process is controlled by mechanosensing osteocytes \cite{Huangetal_FASEBJ_2010,Jacobsetal_ARBE_2010,Roblingetal_ARBE_2006,Huiskesetal_Nature_2000} and provides an example of a homeostatic system where external mechanical loads control the bone mass and structure. Various experimental and theoretical studies analyze the remodeling process on different levels of detail \cite{Dunlopetal_CTI_2009,Fratzletal_PMS_2007,Ruimermanetal_ABME_2005,Tezukaetal_JBMM_2005,Weinkameretal_PRL_2004}. We here introduce a continuous modeling approach with the potential to combine different scales in an efficient way. In an abstract setting we consider bone as a shape changing structure, with concentrations of mechanosensing cells within the bone and resorbing and depositing cells on the bone surface. Depending on the surface concentrations the structure is adapted. In contrast to previous modeling approaches, using micro finite element analysis \cite{Adachietal_JBE_2001,McDonnelletal_JB_2009,Ruimermanetal_JB_2005,Schulteetal_Bone_2011,Schulteetal_Bone_2013}, we describe the structure implicitly using a time-dependent phase field function. This not only leads to a more accurate model, as the artificial voxel-roughness of the bone surface can be avoided, but also to a drastic reduction of system size and required computing time. Phase field models have been developed to describe shape changing structures, e.g. in solidification \cite{ProvatasElder_Wiley} or multiphase fluids \cite{Mauri_Springer}, or have been used as a general numerical tool to solve problems in complex time-evolving geometries \cite{Lietal_CMS_2009} and are today well established in various fields. They also have already been used to compute the mechanical properties of trabecular bone structures \cite{Alandetal_CMBBE_2014} and have been validated against micro finite element analysis. The required implicit description of the bone structure can be obtained from imaging tools, such as $\mu$CT and $\mu$MR by standard algorithms. From the voxel representation of the segmented image a smooth surface triangulation can be constructed, from which a signed distance function and the initial phase field function can be computed. The computation of the remodeling process can then be done on a simple cubic domain, which can easily be decomposed for efficient parallel computations. Fig. \ref{fig1} shows an example of an implicitly described trabecular structure, the phase field representation and details of the phase field function.

The paper is organized as follows: In Section \ref{sec2} we review the current understanding of the bone remodeling process and motivate approximations for our computational approach. Our goal is to introduce a minimal model to demonstrate the potential of the phase field modeling and to point out possible model extensions. We also compare our assumptions with existing micro finite element analysis. In Section \ref{sec3} we propose our mathematical model, the numerical approach and required preprocessing steps and provide details on the $\mu$CT data. In Section \ref{sec4} we describe results which relate the bone morphology to an applied force. In Section \ref{sec6} we discuss the results and give conclusions.

\section{Materials and methods}
\label{sec2}

We briefly review the role of the different cells which are involved in the remodeling process and describe the mechanical properties of bone on the level required for our continuous modeling approach.

\subsection{Osteocytes}

Osteocytes form a network throughout the bone matrix by connecting with each other and the lining cells on the bone surface. They sense mechanical loads and transduce the mechanical signal into a chemical response, which is realized by signaling molecules. On the bone surface these molecules orchestrate the recruitment and activity of osteoblasts and osteoclasts, resulting in the adaptation of bone mass and structure. How the osteocytes sense the mechanical loads and coordinate adaptive alterations in bone mass and architecture is not yet completely understood. For a current review see \cite{Klein-Nulendetal_Bone_2013}. For the mechanical sensing several mechanisms have been proposed \cite{Cowingetal_JBiomechEng_1991}. Within our modeling approach we consider the volumetric compression and the strain energy density as two options to stimulate the osteocytes. The signaling molecules in the bone are then modeled through a diffusion process, with a decay rate and the mechanical stimulus as a source term. This differs from typical micro finite element analysis approaches, where an exponential decay of the signal molecules is assumed and only contributions within a certain distance are accounted for, e.g. \cite{Huiskesetal_Nature_2000,Ruimermanetal_ABME_2005}. 

\subsection{Osteoblasts and osteoclasts}

A large number of hypotheses have been postulated regarding bone-cell communication and the role played by various receptor-ligand pathways. Various modeling approaches
are concerned with the biochemical signaling between active osteoclasts and osteoblasts \cite{Komarovaetal_Bone_2003,Lemaireetal_JTB_2004,Pivonkaetal_Bone_2008}. However, they all work on spatial averages and are not yet incorporated in a computational bone remodeling process. We therefore here only use an effective modeling approach by directly considering the concentration of the signaling molecules at the bone surface as an indicator for resorption and deposition. Similar approximations are made in typical micro finite element analysis approaches with various functional forms. The influences of different remodeling rules like linear, step functions or the originally proposed profile by Frost \cite{Frost_AR_1987} with a lazy zone have been compared in \cite{Dunlopetal_CTI_2009}. While the exact form of the remodeling rule is of importance for quantitative predictions, all forms lead to the emergence of trabecular-like patterns. 

\subsection{Trabecular bone structure and mechanical behaviour}

Also from a mechanical perspective, trabecular bone is a highly complex material, being anisotropic with different strengths in tension, compression, and shear and with 
mechanical properties that vary widely across anatomic sites, and with aging and disease \cite{Oftadehetal_JBE_2015}. Various of these material properties remain uncertain. However, different experiments have shown a linear behavior \cite{Keavenyetal_JB_1994} and also have related the anisotropy of the mechanical properties to the bone structure \cite{Laddetal_JOR_1998}. We therefore also for this part of the model consider the simplest possible approach, an isotropic material with prescribed elastic modulus and Poisson ratio but resolve the trabecular bone structure. The same approximations are made in most micro finite element analysis approaches \cite{Levchuketal_CB_2014,Podshivalovetal_Bone_2011,Arbenzetal_IJNME_2008}.

\section{Theory}
\label{sec3}

\subsection{Mathematical model}
\label{mathmodel}
Let $\Omega_\text{bone}$ be the bone structure and $\Omega$ a cuboid computational domain of size ${\bf L}=(L_x,L_y,L_z)$ such that $\Omega_\text{bone}\subset\Omega$. The mechanical material properties of the trabeculae are supposed to be isotropic with Young's modulus $E$ and Poisson's ratio $\nu$. The bone deformation is described by the displacement vector ${\bf u}$ obeying the partial differential equation 
\begin{eqnarray}
\label{eq_sigma}
\nabla\cdot {\bf \sigma} = 0 \qquad\mbox{in }\Omega_\text{bone} \label{eq1}
\end{eqnarray}
with the linear elastic stress tensor
\begin{eqnarray}
{\bf \sigma} = \mu(\nabla{\bf u}+\nabla{\bf u}^T)+\lambda I \nabla\cdot {\bf u} \label{eq2}
\end{eqnarray}
where $I$ is the identity matrix and $\mu$, $\lambda$ are the Lame coefficients 
$$
\mu = \frac{E}{2 (1 + \nu)}, \quad \lambda = \frac{E \nu}{(1 + \nu)(1- 2 \nu)}.
$$
Compression is achieved by applying Dirichlet boundary conditions to the normal component of ${\bf u}$ on $\partial \Omega$, such that 
$u_{x,y,z}= 0$ if $x,y,z=0$ and $u_{x,y,z}= \bar{u}_{x,y,z}$ if $x=L_x, y = L_y, z = L_z$, while the tangential displacement components are free and $\bar{u}_{x,y,z}$ are adapted such that the normal forces $(F_x, F_y, F_z)$ are kept constant.  At the remaining boundaries $\partial \Omega_\text{bone}$ we assume no outer forces and set ${\bf \sigma} \cdot {\bf n} = 0$, where ${\bf n}$ is the normal to $\Omega_{bone}$.

We assume that osteocytes are continuously distributed through $\Omega_\text{bone}$ and that a growth stimulating signal is produced in the osteocytes stimulated by either the volumetric compression or the strain energy density 
\begin{align}
\label{eq_stimuli}
S_{\rm{vc}} &= \left| \nabla\cdot{\bf u} \right|  \;\;&\text{(volumetric compression)} \\
S_{\rm{sed}}   &= \frac{1}{4} {\bf \sigma}:(\nabla{\bf u}+\nabla{\bf u}^T)  \;\;&\text{(strain energy density)} 
\end{align}
The signal is propagated by a diffusive process with a constant decay rate. Denoting the concentration of the signaling molecule by $c$, this leads to the equation 
\begin{align}
k c - d\Delta c &= S & \text{in}~\Omega_\text{bone} \label{c eq1}
\end{align}
with boundary condition ${\bf n}\cdot\nabla c = 0$ on $\partial\Omega_\text{bone}$, where $k$ is the decay rate and $d$ the diffusion coefficient, both assumed to be constant and $S = S_{\rm{vc,sed}}$.
Due to the different time scale, if compared with the remodeling process, we consider the stationary solution.

Finally growth is triggered directly according to the concentration $c$ at the trabecular surface.
We assume that the growth velocity in normal direction $V = V_{\rm{lin,lazy}}$ depends linearly on the signal strength, without or with a lazy zone
\begin{align} 
V_{\rm{lin}} &= \alpha c - \beta & \text{on}~\partial\Omega_\text{bone} \label{V eq1} \\
V_{\rm{lazy}} &= \alpha (c-min(T, max(-T, c-1)))-\beta & \text{on}~\partial\Omega_\text{bone} \label{V eq2}
\end{align}
with positive constants $\alpha$ and $\beta$ and an intermediate range for $c$ where no growth occurs, controlled by the threshold value $T$.

Mathematically the resulting system of equations (\ref{eq1}) - (\ref{V eq2}) is a free boundary problem. According to the proposed normal forces $(F_x,F_y,F_z)$ on $\delta \Omega$ the displacement $\mathbf{u}$ and the concentration $c$ have to be computed in $\Omega_\text{bone}$. The obtained signal $c$ at $\delta \Omega_\text{bone}$ is then used to update the time dependent bone structure $\Omega_\text{bone}$. This non-linear problem has to be iterated until the solution for 
$\mathbf{u}, c$ and $\Omega_\text{bone}$ converge.

\subsection{Numerical solution}

In micro finite element analysis the time dependent bone structure is accounted for by adding and removing voxels to $\Omega_\text{bone}$. This is not only costly as it requires a high spatial resolution, it also leads to an artificial roughness of the bone surface. Various numerical methods have been proposed to avoid such manipulation of the computational domain. One of the most successful approaches is the phase field or diffuse domain approach \cite{Lietal_CMS_2009}. Here $\Omega_\text{bone}$ is described implicitly through a smooth phase field function $\phi = 0.5 (1 - \tanh(r/(\sqrt{8} \epsilon)))$ in $\Omega$, with a small parameter $\epsilon$ determining the width of the diffuse interface and the signed distance function $r$, with $r = 0$ at $\delta \Omega_\text{bone}$, $r > 0$ in $\Omega_\text{bone}$ and $r < 0$ in $\Omega \setminus \Omega_\text{bone}$.  For $\epsilon \to 0$, $\phi$ converges to the characteristic function for $\Omega_\text{bone}$. Using $\phi$ we can now reformulate the problem in the time-independent domain $\Omega$. The diffuse domain approximation of eqs. (\ref{eq1}) and (\ref{eq2}) reads  
\begin{equation}\label{eq1_phi}
 \mu  \nabla \cdot \left[\phi (\nabla \mathbf{u} + \nabla \mathbf{u}^T)\right] + \lambda \nabla \left[\phi ~\nabla \cdot \mathbf{u}\right] = 0 \quad\mbox{in} ~\Omega.
\end{equation}
The boundary condition $\sigma \cdot \mathbf{n} = 0$ is implicitly included, see \cite{Alandetal_CMBBE_2014}. For eq. (\ref{c eq1}) we obtain
\begin{equation}\label{eq2_phi}
k \phi c - d \nabla \cdot (\phi \nabla c) = \phi S \quad\mbox{in} ~\Omega,
\end{equation}
which follows along the same arguments and also already includes the boundary condition $\mathbf{n}\cdot \nabla c = 0$. 
Adapting the bone domain with normal velocity $V$ can be realized by solving 
\begin{equation}\label{eq3_phi}
\partial_t \phi + V |\nabla \phi| = \gamma (- \phi^3 + 1.5 \phi^2 - 0.5 \phi) + \gamma \epsilon^2 \frac{\nabla \phi^T \cdot \nabla \nabla \phi \cdot \nabla \phi}{|\nabla \phi|^2} \quad\mbox{in} ~\Omega
\end{equation}
with a mobility factor $\gamma$, see \cite{Folchetal_PRE_1999}. The terms on the right hand side essentially guarantee the $\tanh$-profile of $\phi$ and the left hand side models the transport of $\phi$ by $V$. Using matched asymptotic analysis one can show, see \cite{Lietal_CMS_2009}, that the system of equations (\ref{eq1_phi}) - (\ref{eq3_phi}) converge for $\epsilon \to 0$ to the original problem (\ref{eq1}) - (\ref{c eq1}) with the boundary moved with velocity $V$ and boundary conditions $\sigma \cdot \mathbf{n} = 0$ and $\nabla c \cdot \mathbf{n} = 0$. We can now use standard discretization techniques to solve for $\mathbf{u}, c$ and $\phi$ in $\Omega$, again iteratively until the system converges.  

For a computationally efficient treatment adaptive mesh refinement is used. A high resolution within the diffuse interface is required, which needs to be of order $\epsilon$ and is comparable to the initial voxel scale. However, away from the diffuse interface a coarser mesh can be used, which drastically reduces the computational cost. The efficiency can further be improved by using differently refined meshes for the components \cite{Voigtetal_JCS_2011,Lingetal_CAMS_2016}. We here use a parallel adaptive finite element approach on unstructured meshes which is implemented in the open source toolbox AMDiS \cite{Veyetal_CMS_2010,Witkowskietal_AMS_2015}. 

\subsection{Preprocessing}

\begin{figure}[h]
\centering
\includegraphics[width=.49\linewidth]{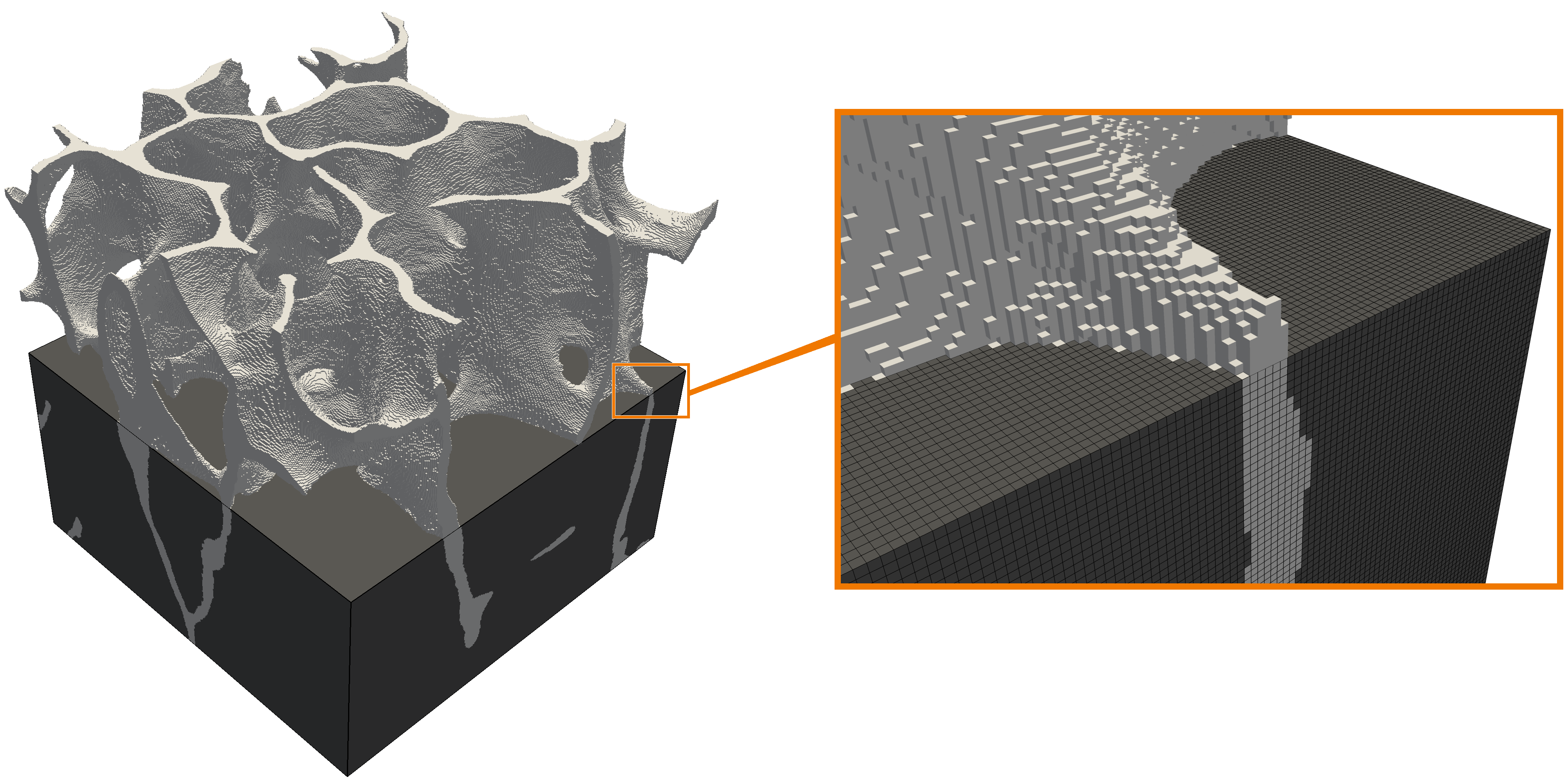}
\includegraphics[width=.49\linewidth]{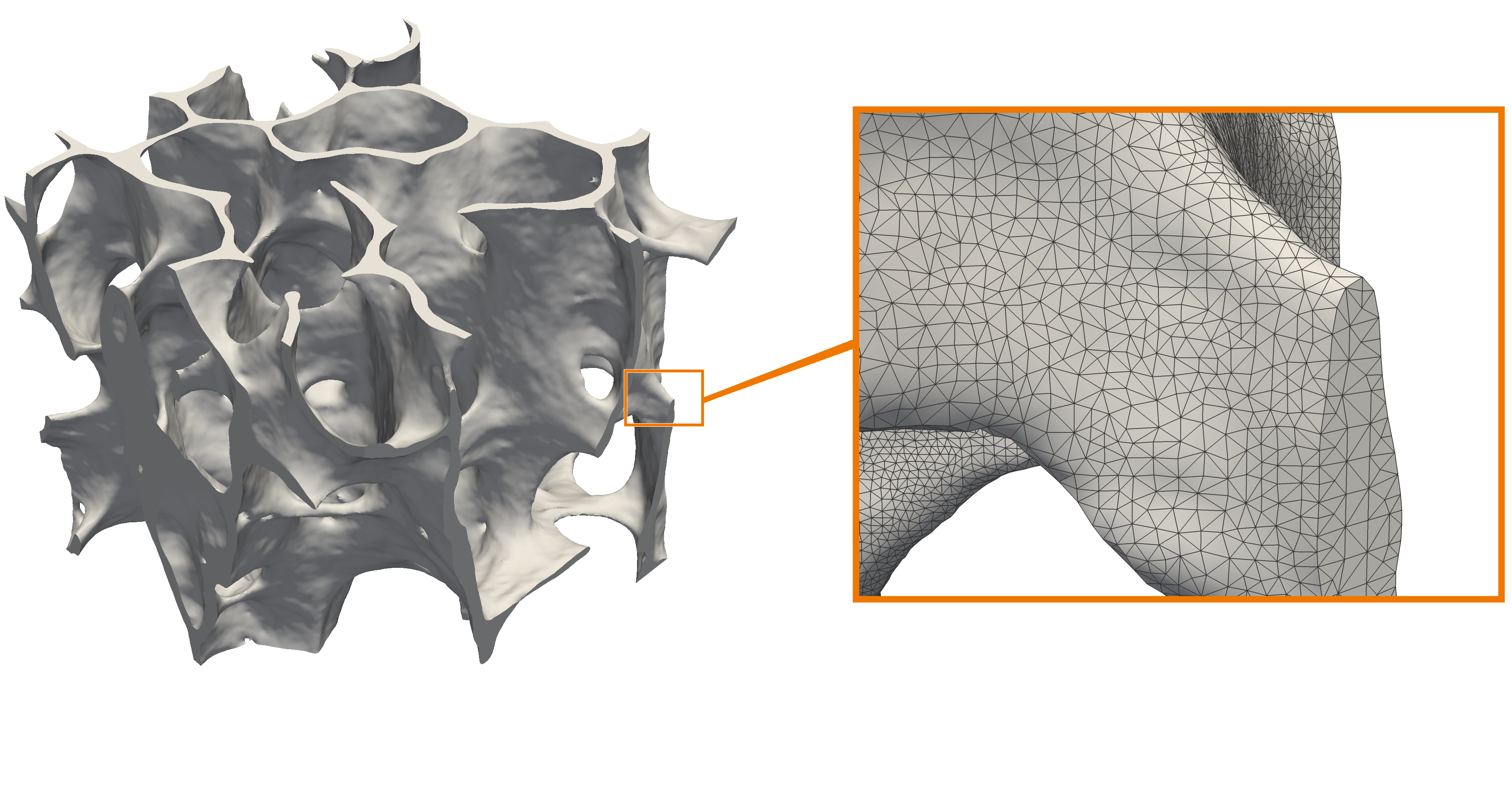}
\includegraphics[width=.45\linewidth]{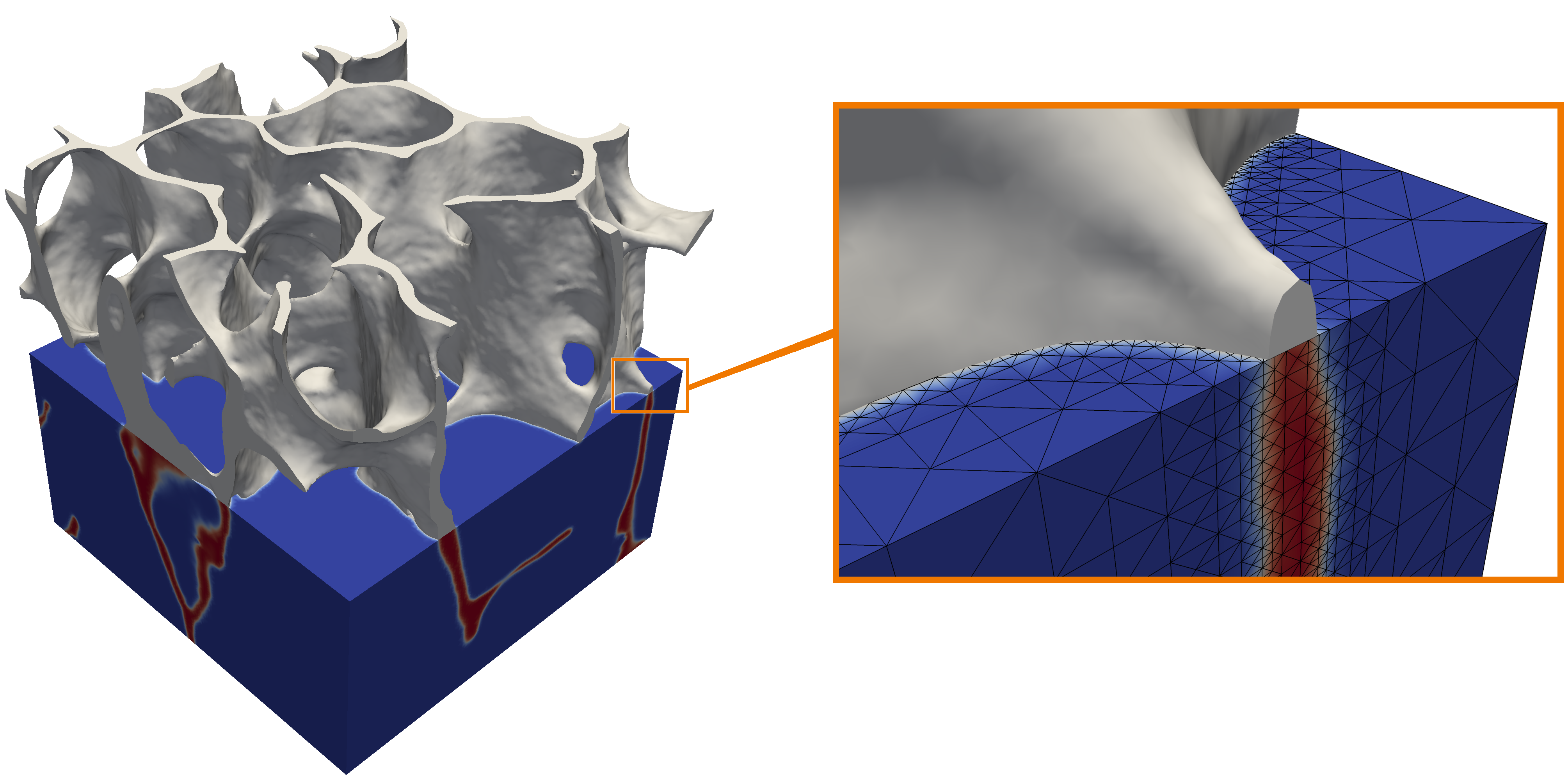}
\caption{Cuboidal segment of a sheep bone. From top left to bottom right: Segmented voxel representation (bone in light gray), triangulated smooth surface representation and phase field description with smooth level surface and adapted volume triangulation.}
\label{fig9}
\end{figure}

To use the phase field approach for real bone structures requires various preprocessing steps. From the segmented data on the voxel scale, first a surface triangulation is constructed, from which in a second step the signed distance and phase field function can be computed. Fig. \ref{fig9} shows the three steps. Various approaches are available for achieving these steps. We here use ParaView (http://www.paraview.org) for the generation and smoothing of the surface mesh. Mesh generation has become a standard tool. However, the quality of these meshes in terms of regularity is often very poor. Automatic construction of high quality surface meshes is still not possible for arbitrary surfaces and is an ongoing research topic. The available algorithms in ParaView provide a reasonable compromise between usability and mesh quality. To compute the signed distance function we embed the structure in a cube and adaptively refine the mesh until we obtain the proposed accuracy to resolve the surface. For each node we compute the shortest distance to the surface using raytracing. The approach is implemented in MeshConv (https://gitlab.math.tu-dresden.de/wir/meshconv) and produces the initial mesh and signed distance function for the computation in AMDiS (https://gitlab.math.tu-dresden.de/wir/amdis). All these software tools are available under an open source license. 

\subsection{Validation}

Before we simulate bone remodeling on real bone structures, we consider two simple examples, a cylinder and a cross, for which we compare the results with a micro finite element analysis approach \cite{Huiskesetal_Nature_2000,Ruimermanetal_ABME_2005}, and in the cylinder case also with a semi-analytical solution. We consider both types of stimuli, the volumetric compression and the strain energy density and consider the linear growth law eq. (\ref{V eq1}).

For a cylinder with radius $R$ and a compression force $F_z$, we obtain for the stimuli $S_{\rm{vc}}= (1-2\nu) \frac{F_z}{E} \frac{1}{\pi R^2}$ and $S_{\rm{sed}}  = \frac{1}{2} \frac{F_z^2}{E} \frac{1}{\pi^2 R^4}$. Both are constant in $\Omega_\text{bone}$, which allows to solve eq.~\eqref{c eq1}. We obtain $c = \frac{S}{k}$ and thus from eq. \eqref{V eq1} the growth law $V_{\rm{lin}} = \frac{\alpha}{k}S - \beta$ from which the evolution of the cylinder can be obtained.

The micro finite element analysis is based on \cite{Huiskesetal_Nature_2000,Ruimermanetal_ABME_2005}. The domain $\Omega$ is divided into a regular voxel mesh with mesh size $h$. Each voxel has a continuous value $g \in \left[0,1\right]$, where bone is associated with $g = 1$ and bone marrow corresponds to $g = 0$. Transient states correspond to voxels which are partly bone. The elastic modulus for a voxel and the stimuli are defined as $gE$ and $gS$, respectively. The Poisson ratio remains independent of $g$. Instead of directly solving for the normal velocity $V$ we first compute the change rate of a voxel's $g$ value at $\partial \Omega_{bone}$ and the neighboring marrow voxels by solving $\partial_t g = \alpha c - \beta$ and restricting $g \in \left[0,1\right]$. The normal velocity then follows by multiplying the change rate with the voxel size $V=h~\partial_t g$ and applying corrections for the lattice anisotropy. For the numerical solution of eqs. (\ref{eq_sigma}) and (\ref{c eq1}), each voxel with $g = 1$ is converted to a hex-8 brick element and the resulting systems are solved by standard finite element analysis.

\begin{figure}
\begin{tabular}{lccc|cccc}
&&growth&&&shrinkage \\
& $t=0$ & $t=0.1$ & $t=2.0$ & $t=0$ & $t=0.1$ & $t=2.0$  \\
\begin{minipage}{1cm}
phase field
\end{minipage} 
&
\includegraphics[width=0.12\textwidth, trim=115px 70px 220px 90px, clip]{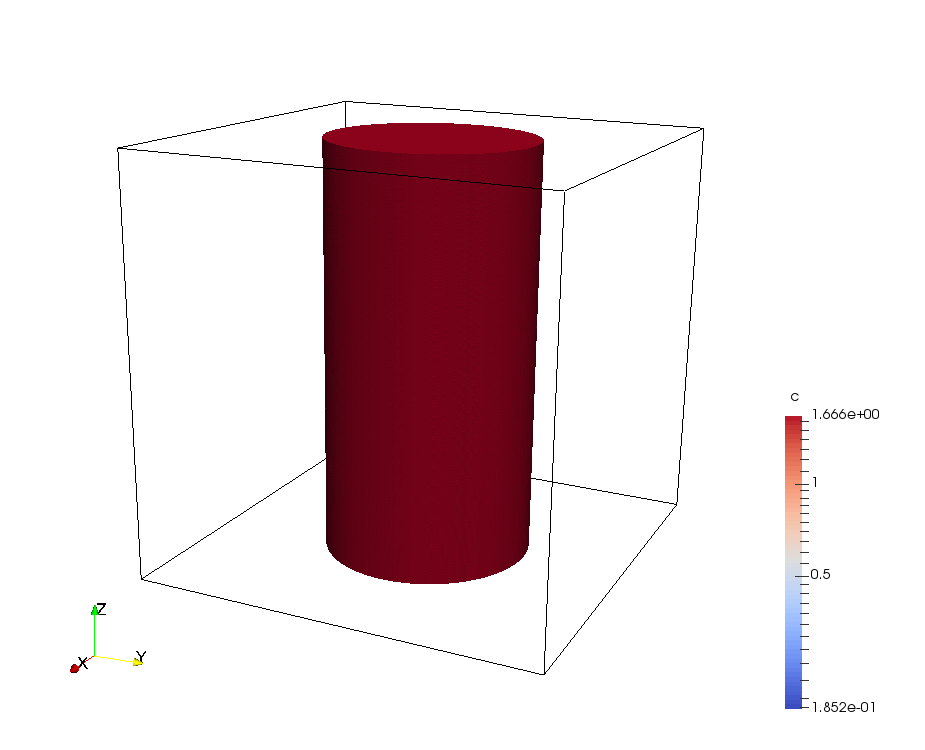} &
\includegraphics[width=0.12\textwidth, trim=115px 70px 220px 90px, clip]{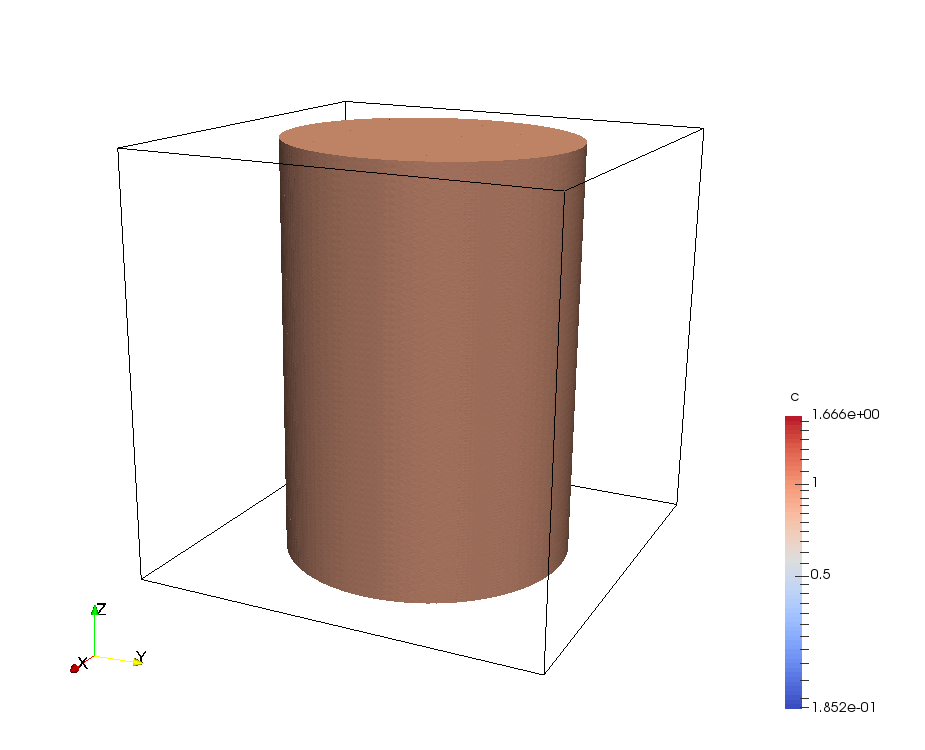} &
\includegraphics[width=0.12\textwidth, trim=115px 70px 220px 90px, clip]{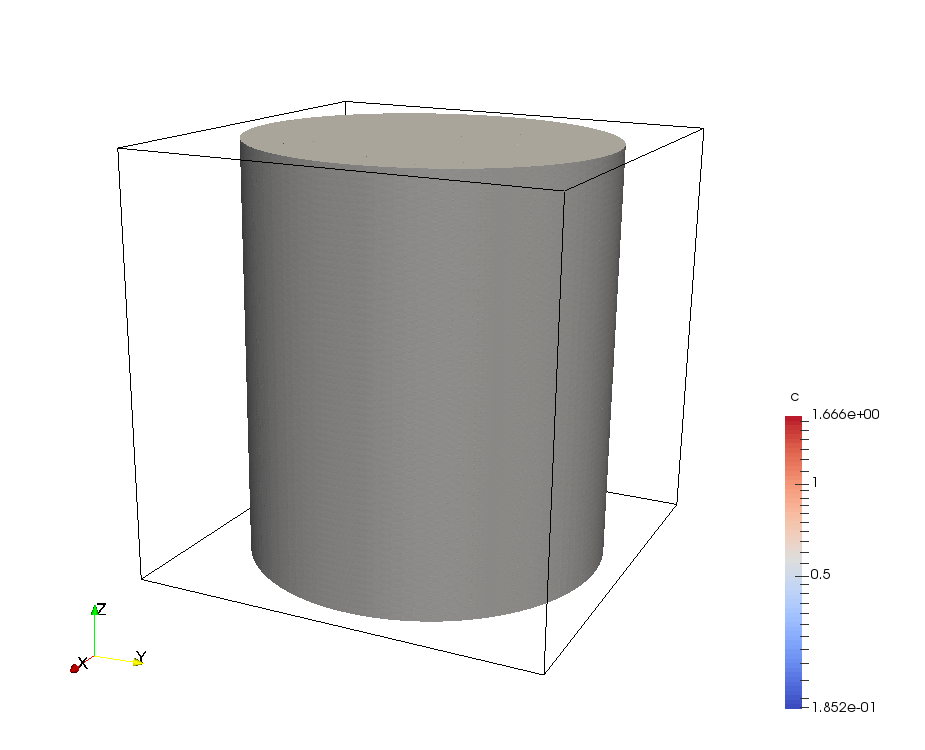} &
\includegraphics[width=0.12\textwidth, trim=115px 70px 220px 90px, clip]{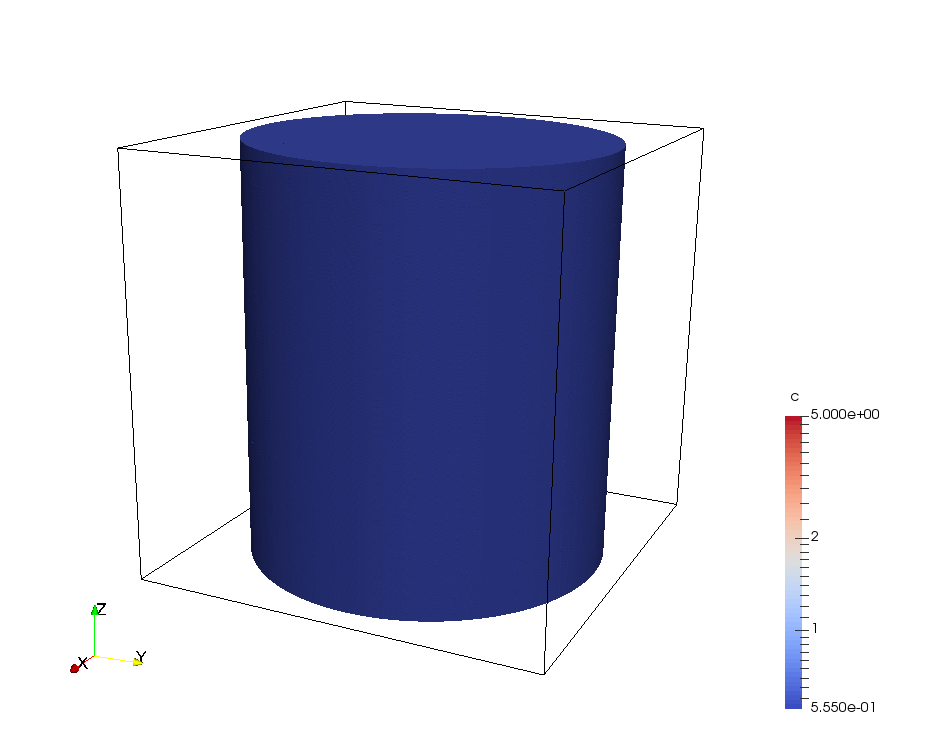} &
\includegraphics[width=0.12\textwidth, trim=115px 70px 220px 90px, clip]{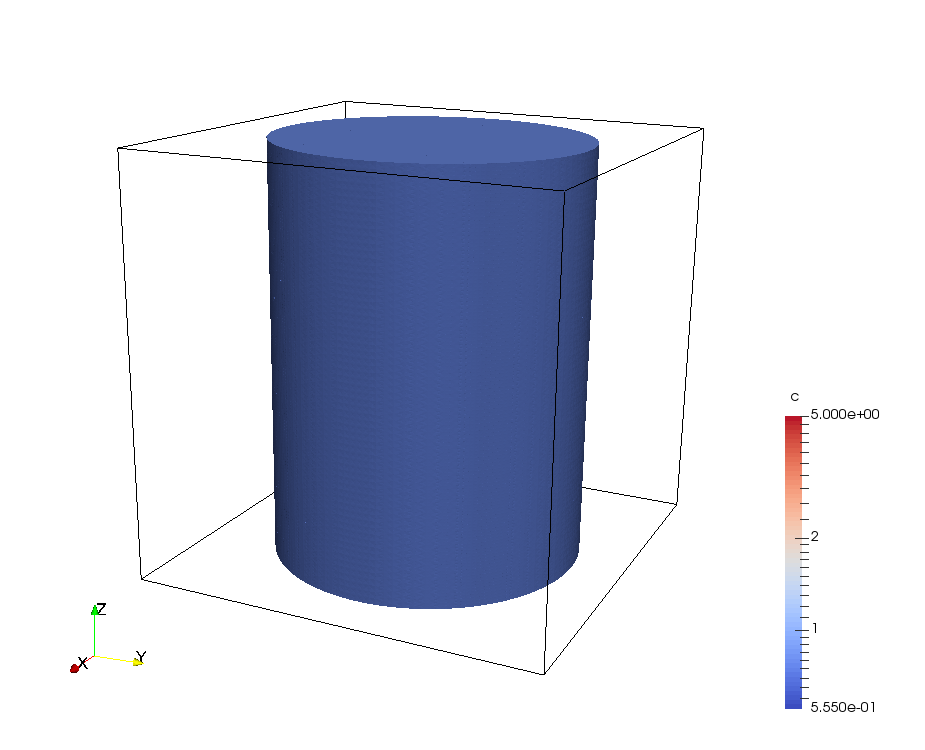} &
\includegraphics[width=0.12\textwidth, trim=115px 70px 220px 90px, clip]{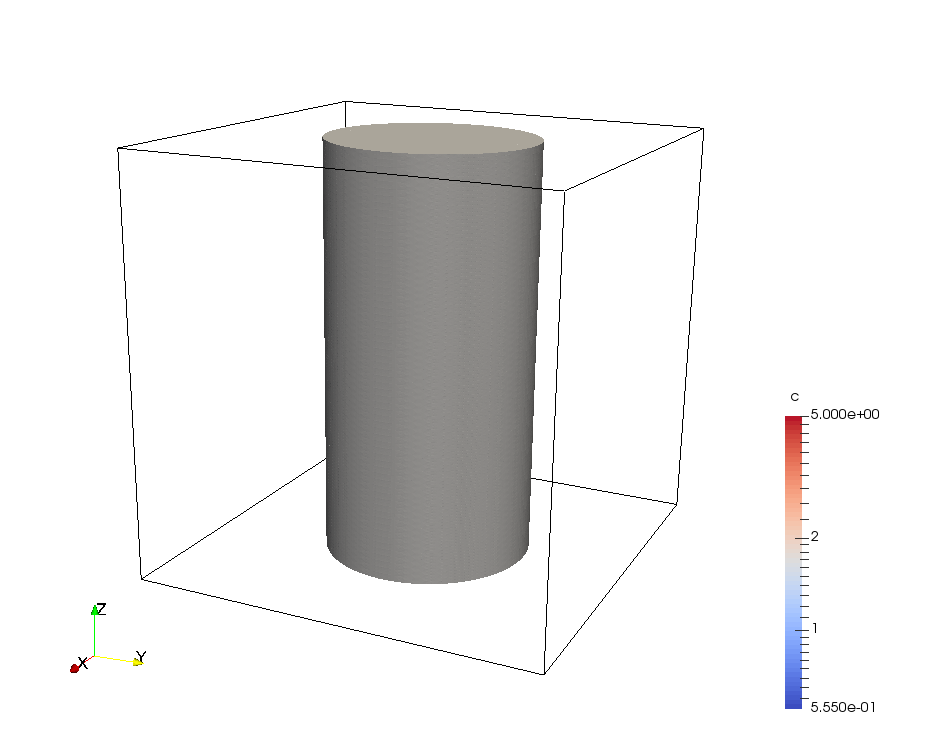} &
\includegraphics[width=0.035\textwidth]{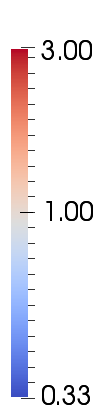} 
\\
\begin{minipage}{1cm}
voxel
\end{minipage} 
&
\includegraphics[width=0.12\textwidth, trim=115px 70px 220px 90px, clip]{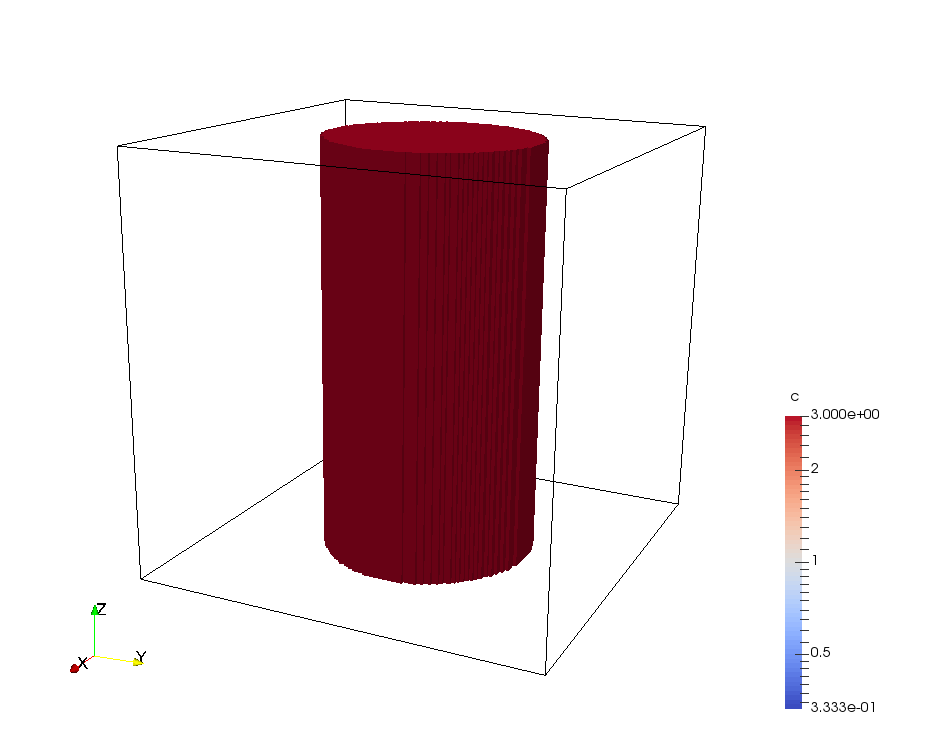} &
\includegraphics[width=0.12\textwidth, trim=115px 70px 220px 90px, clip]{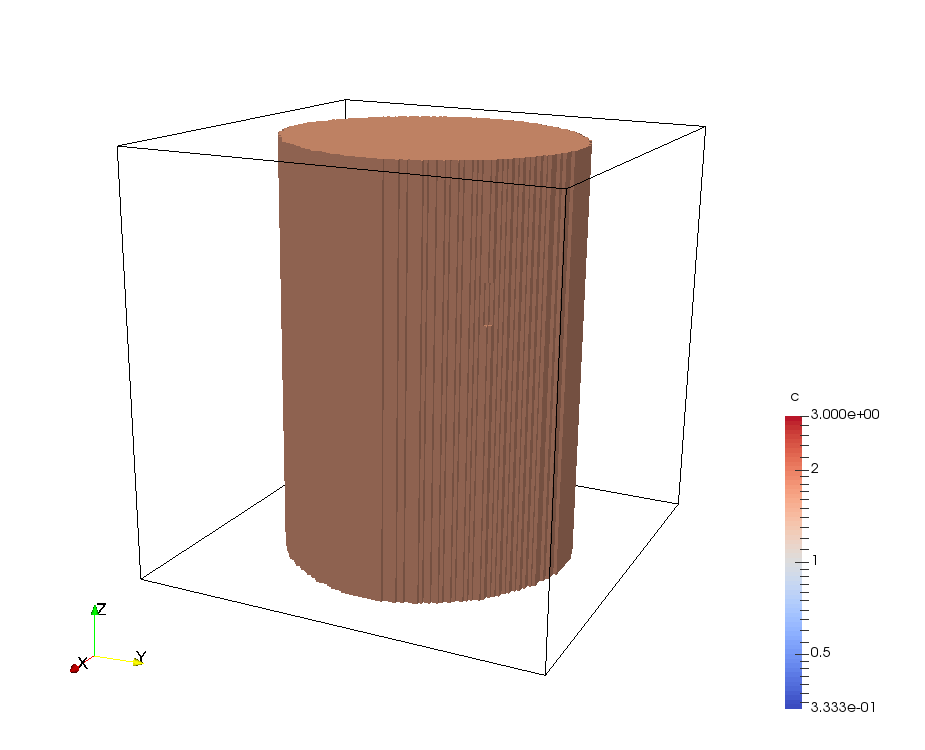} &
\includegraphics[width=0.12\textwidth, trim=115px 70px 220px 90px, clip]{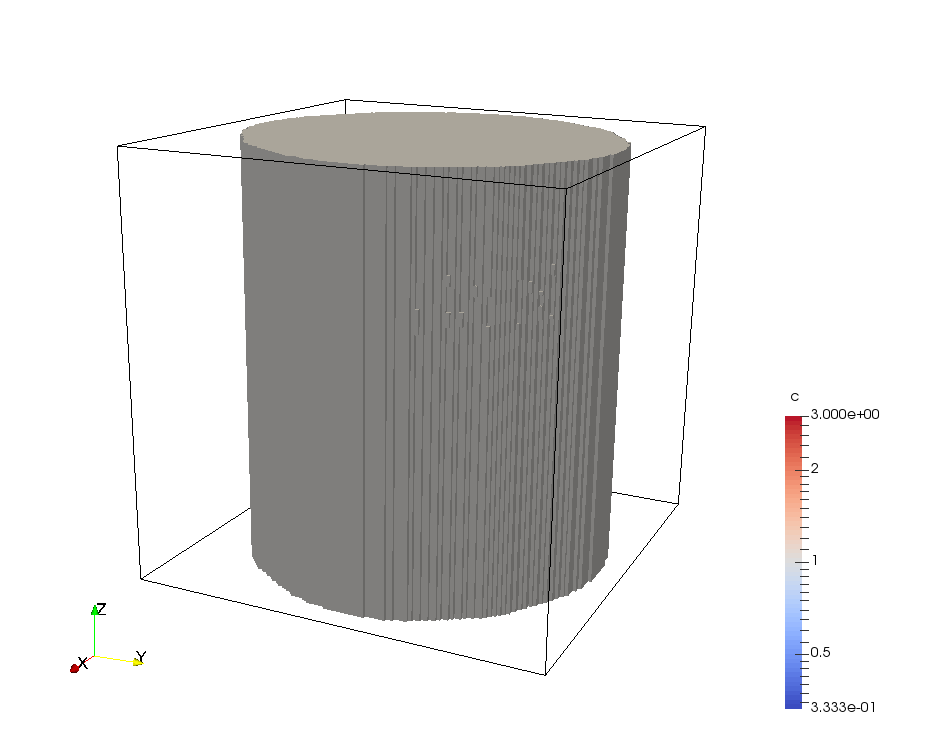} &
\includegraphics[width=0.12\textwidth, trim=115px 70px 220px 90px, clip]{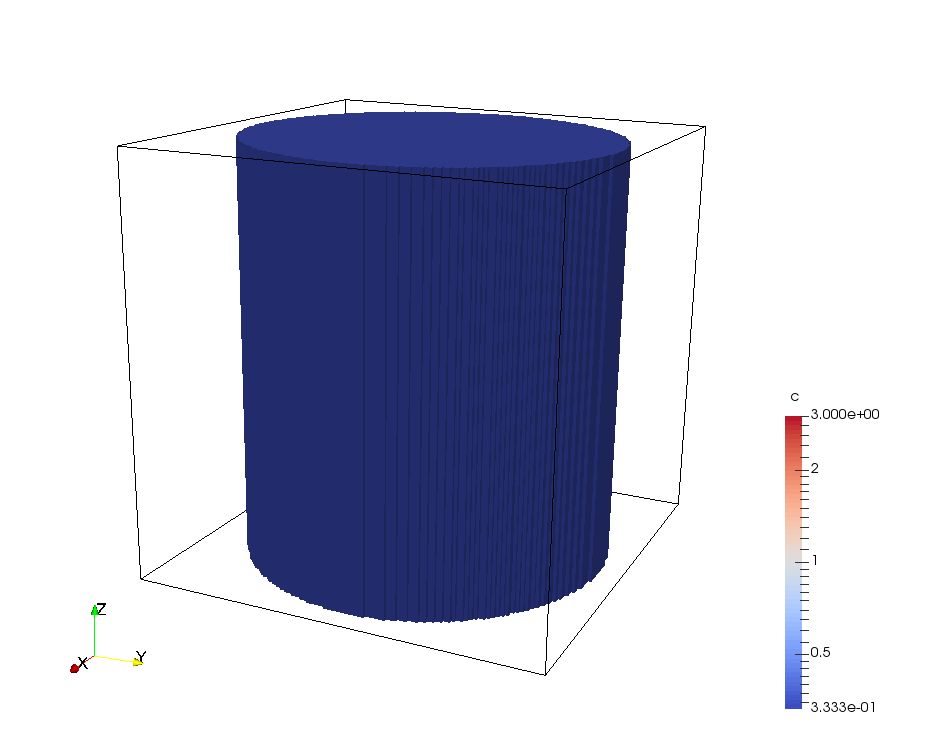} &
\includegraphics[width=0.12\textwidth, trim=115px 70px 220px 90px, clip]{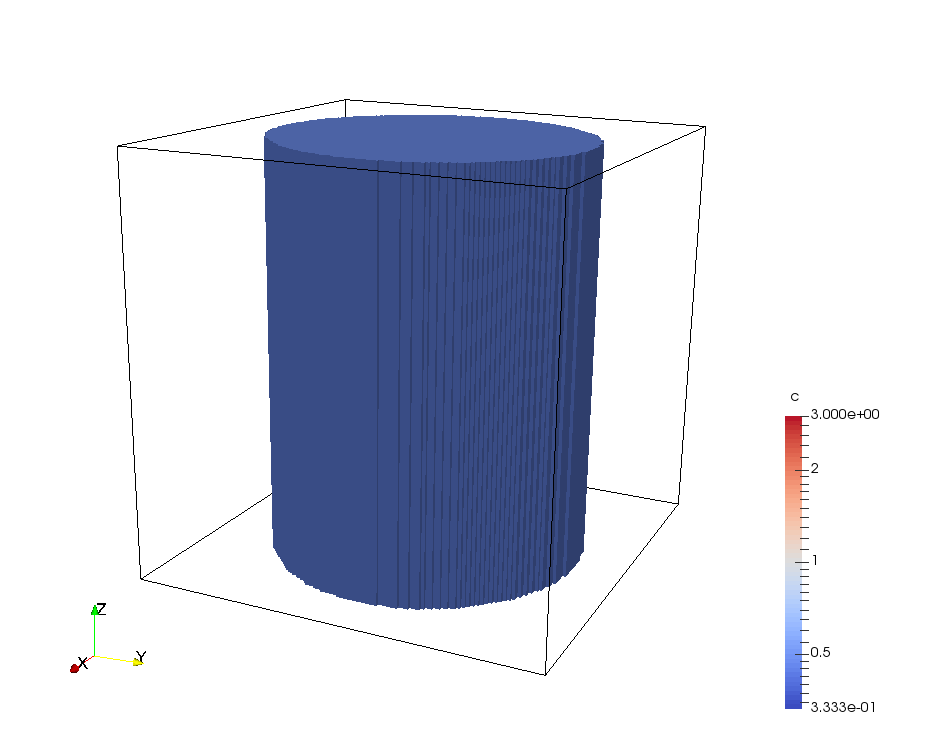} &
\includegraphics[width=0.12\textwidth, trim=115px 70px 220px 90px, clip]{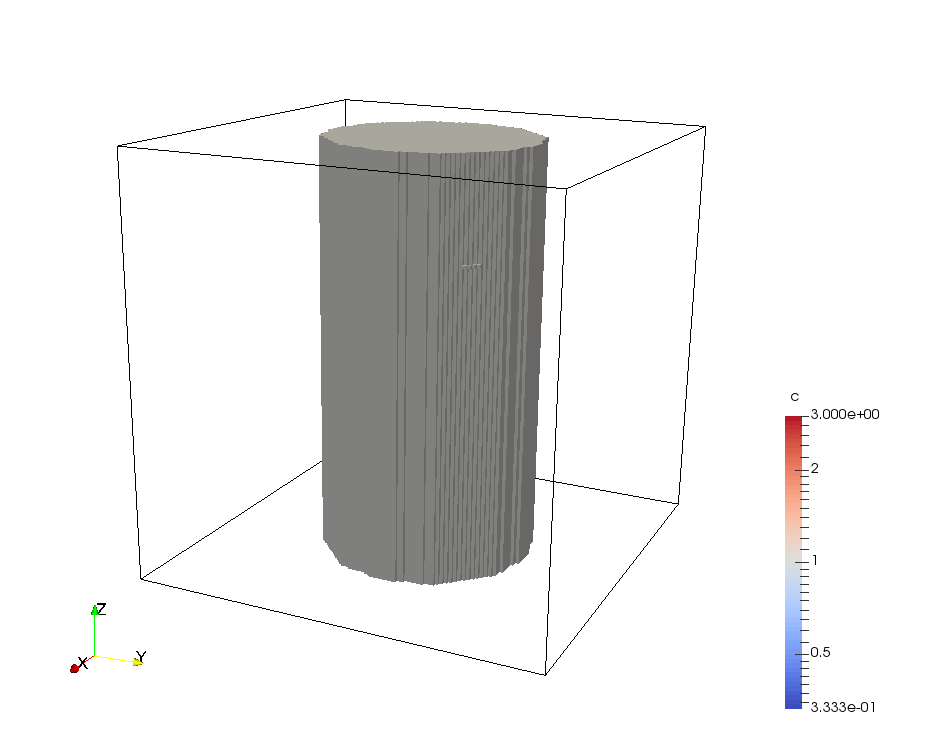} \\
\end{tabular}
\caption{Cylinder growth (left) and shrinkage (right). The images correspond to the use of the volumetric compression as stimulus.
The color coding corresponds to the scaled concentration $\alpha c$.}
\label{fig:cylinder}
\end{figure}

\begin{figure}
\centering
\includegraphics[width=0.49\textwidth]{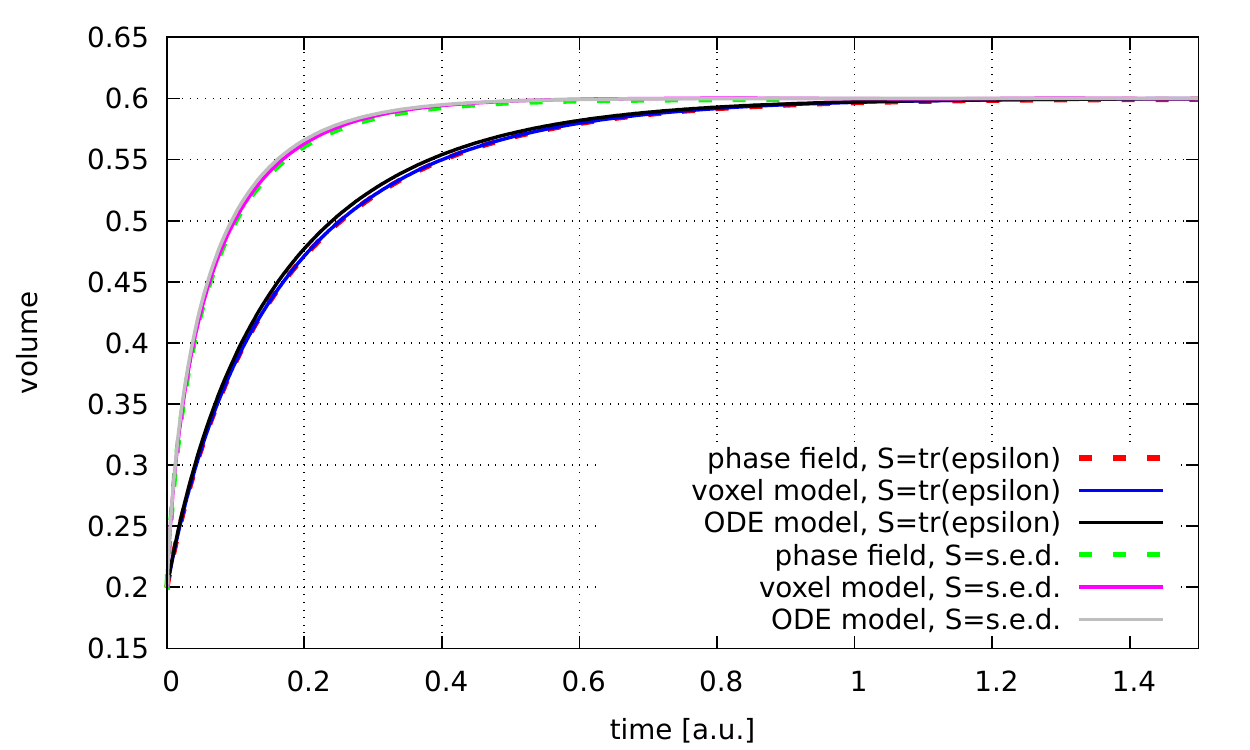}
\includegraphics[width=0.49\textwidth]{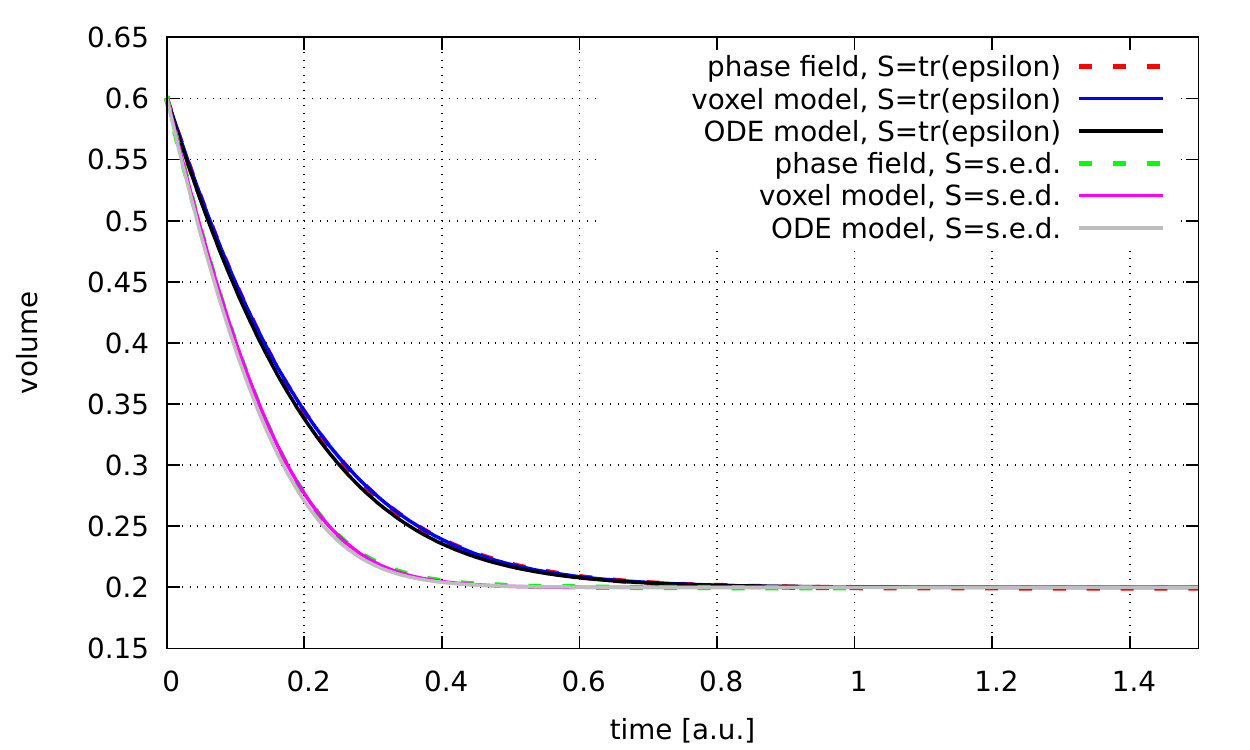}
\caption{Cylinder volume over time for growth (left) and shrinkage (right) for volumetric compression and strain energy density as stimulus. Shown are the phase field and micro finite element results together with the semi-analytic solution (ODE model).}
\label{fig:cylinder graphs}
\end{figure}

\begin{figure}
\begin{tabular}[b]{lcccc|}
& $t=0$ & $t=0.1$ & $t=2.0$ \\
\begin{minipage}{1cm}
phase field
\end{minipage} 
&
\includegraphics[width=0.11\textwidth, trim=115px 70px 220px 90px, clip]{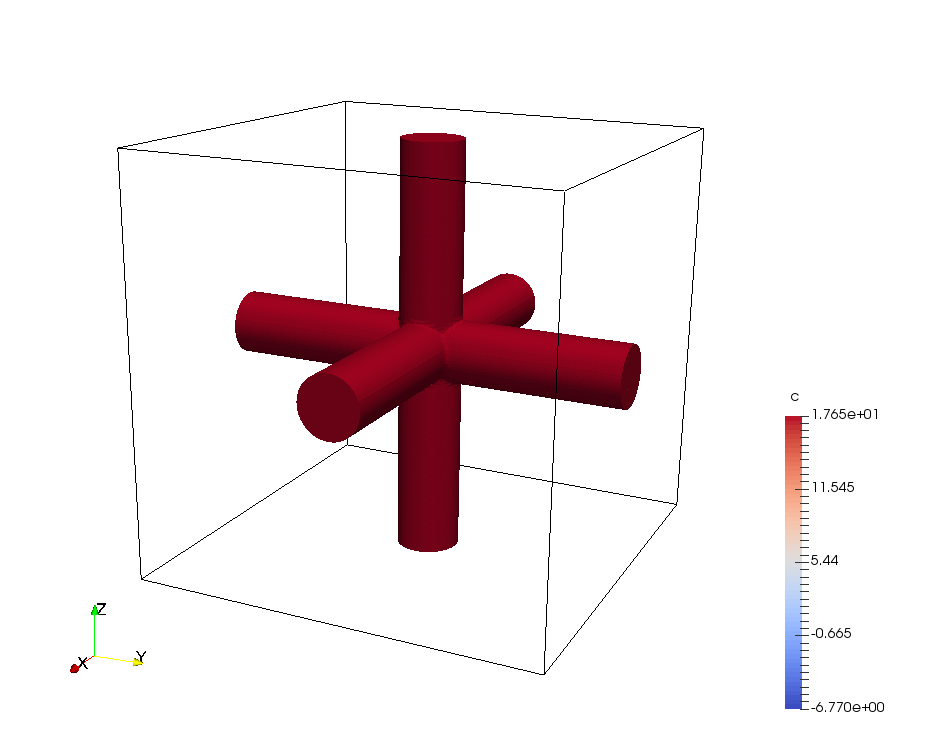} &
\includegraphics[width=0.11\textwidth, trim=115px 70px 220px 90px, clip]{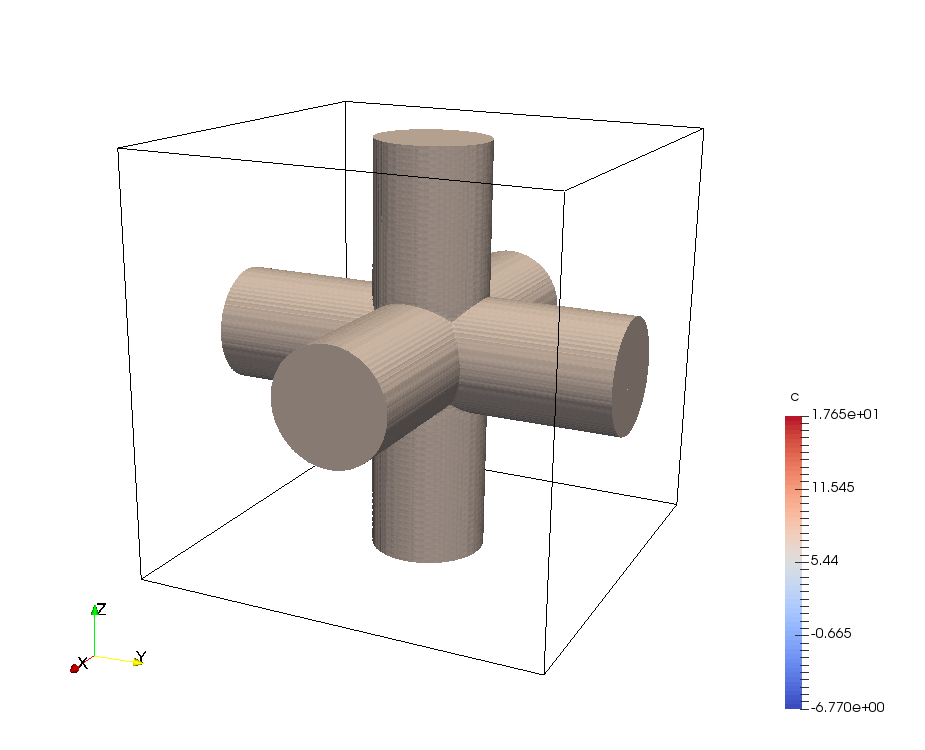} &
\includegraphics[width=0.11\textwidth, trim=115px 70px 220px 90px, clip]{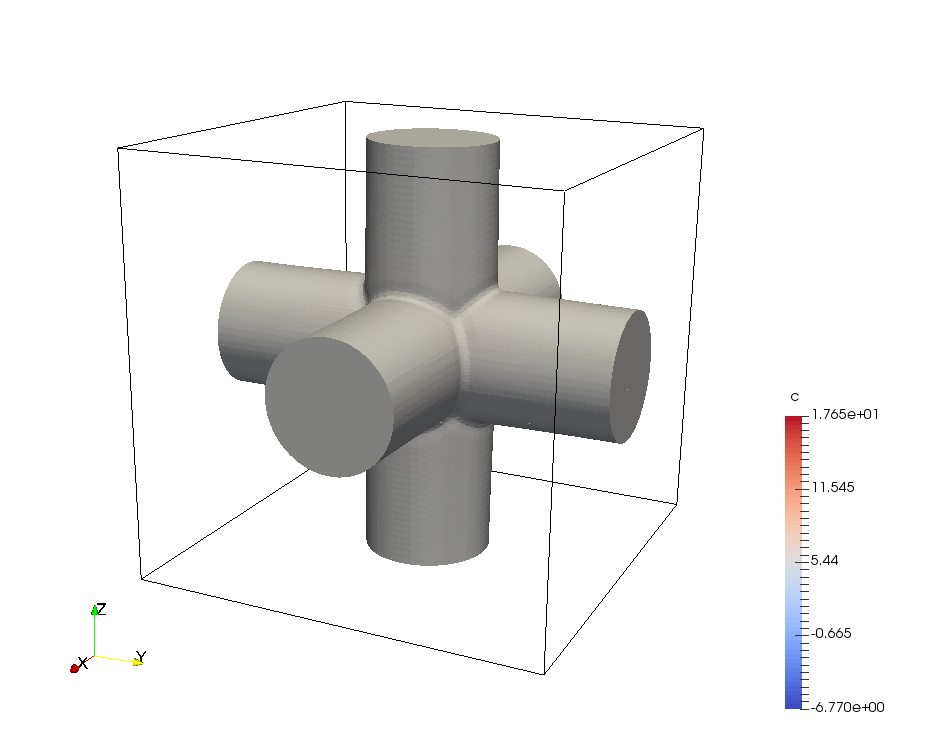} &
\includegraphics[width=0.032\textwidth]{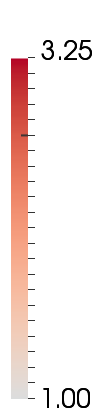} \\
\begin{minipage}{1cm}
voxel
\end{minipage} 
&
\includegraphics[width=0.11\textwidth, trim=115px 70px 220px 90px, clip]{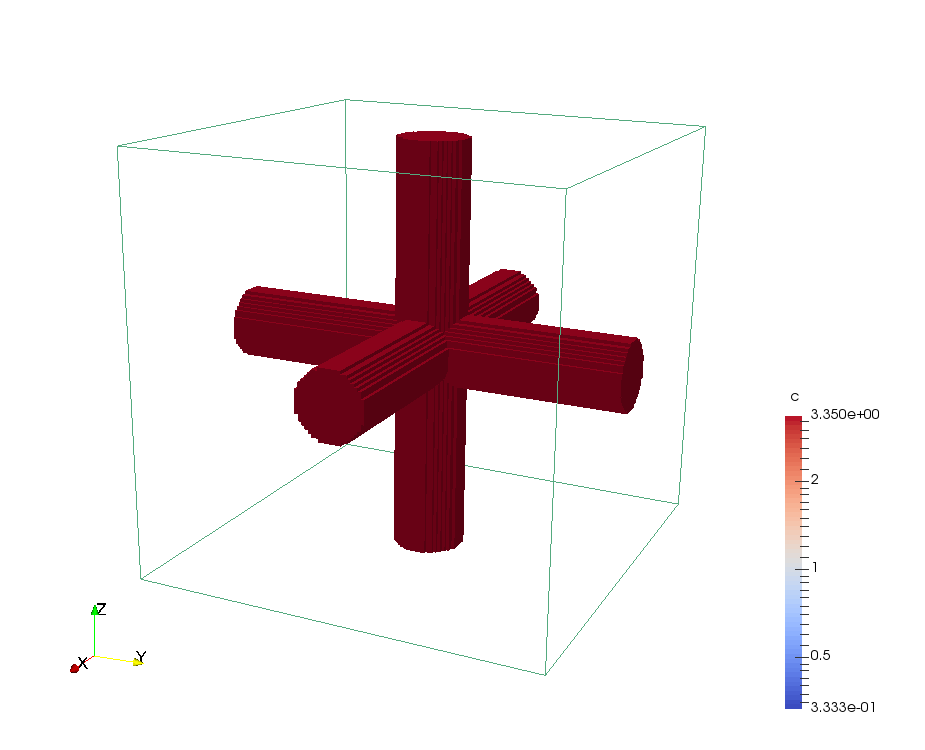} &
\includegraphics[width=0.11\textwidth, trim=115px 70px 220px 90px, clip]{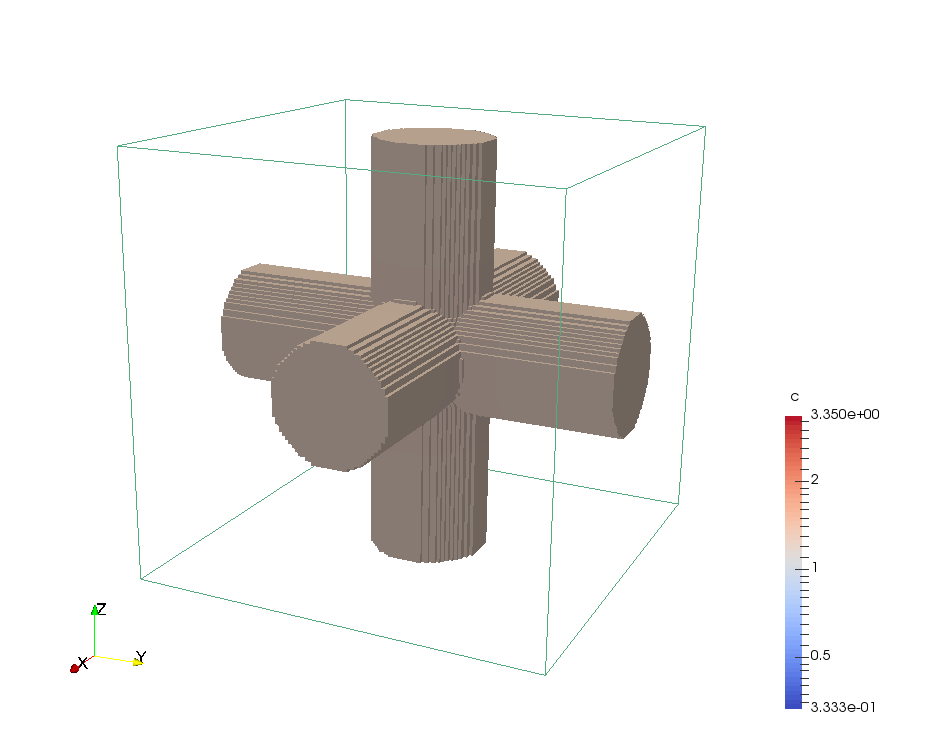} &
\includegraphics[width=0.11\textwidth, trim=115px 70px 220px 90px, clip]{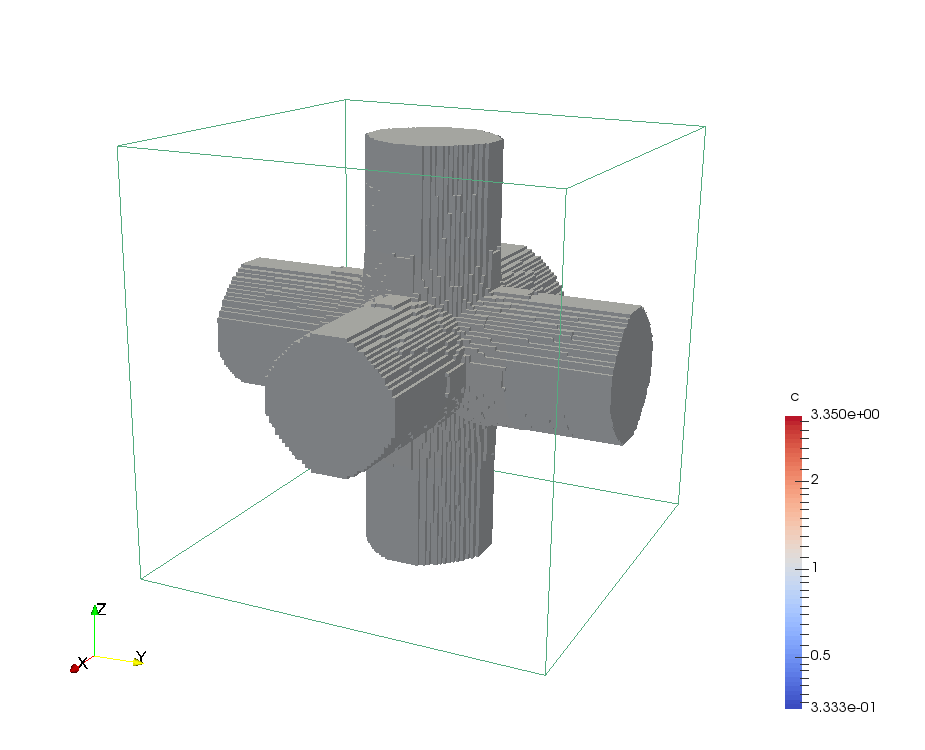} &
 \\
\end{tabular}
\includegraphics[width=0.42\textwidth]{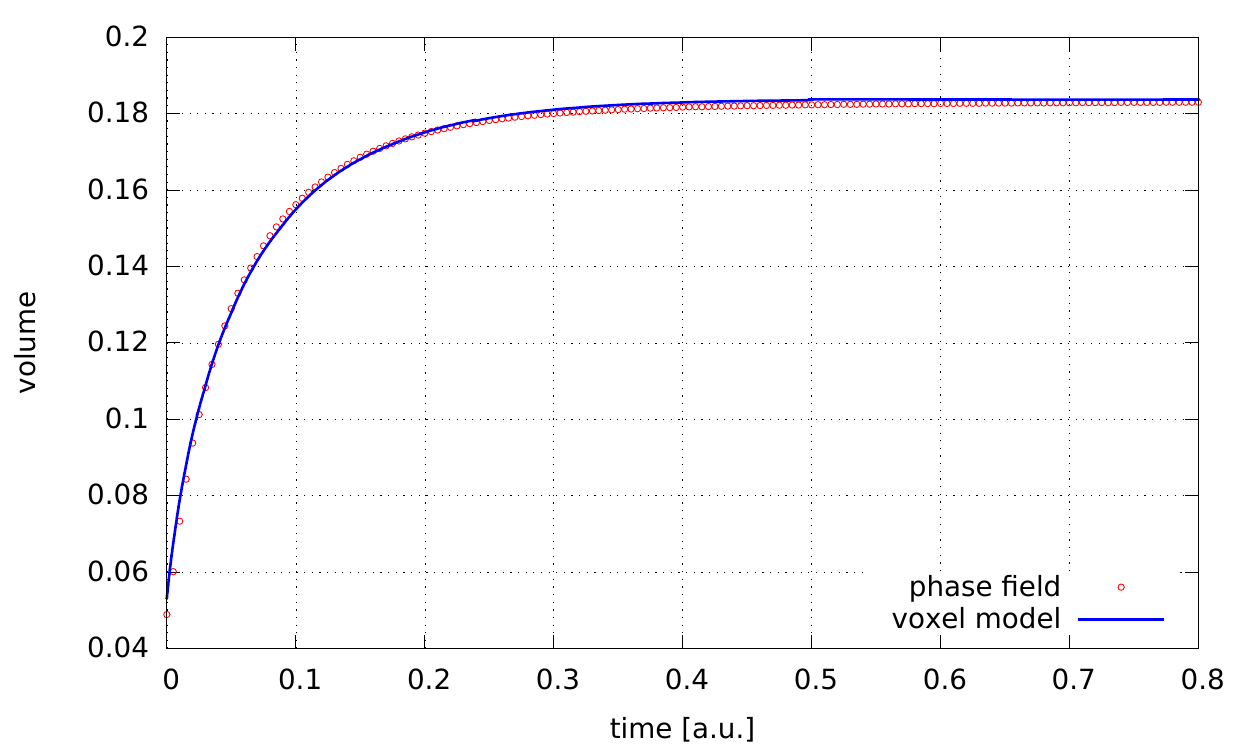}
\caption{Cross growth evolution (left) and cross volume over time (right). The volumetric compression is used as stimulus. The color coding corresponds to the scaled concentration $\alpha c$, where $\alpha c=1$ indicates the approached stationary state.}
\label{fig:kreuz}
\end{figure}

For the comparison we only consider non-dimensional values. The computational domain is the unit cube with $L_x = L_y = L_z = 1$. The voxel size, as well as the minimum grid spacing in the phase field approach is set to $h = 1/128$. Time steps were chosen adaptively. Other parameters are
$d = 1$, $k = 1/2$, $F_{\rm{z}} = 1$ for the cylinder and $F_{\rm{x}}=F_{\rm{y}}=F_{\rm{z}}=1$ for the cross, as well as $\beta=1$ and $\alpha$ depending on the considered stimulus $\alpha_{\rm{sed}}=0.36$ and $\alpha_{\rm{vc}}=0.9$ for the growth experiment and $\alpha_{\rm{sed}}=0.04$ and $\alpha_{\rm{vc}}=0.3$ for the shrinkage simulations. These parameters lead to a stationary state with $R=\sqrt{0.6 / \pi}$ and $R=\sqrt{0.2 / \pi}$ for growth and shrinkage, respectively. As initial condition we thus use the opposed value, i.e. $R=\sqrt{0.2 / \pi}$ (growth), $R=\sqrt{0.6 / \pi}$ (shrinkage). For the simulation of a single growing cross, $\alpha_{\rm{vc}}=0.0919$ is used and the initial radius on each side is set to $R=\sqrt{0.02/ \pi}$. The remaining parameters for the phase field approach are $\epsilon = 0.005$ and $\gamma = 10$. 

In all these cases the solution of the free boundary problem converge to a stationary solution providing the adapted bone morphology to the applied force, see Figs.~\ref{fig:cylinder}-\ref{fig:kreuz}. The results are in excellent agreement with the micro finite element results and the analytic solution, where available.

\subsection{$\mu$CT protocol}

A cylindrical shaped specimen of vertebra L1 was scanned in a $\mu$CT system SkyScan 1173 (Bruker MicroCT, Kontich, Belgium). Details of the scanning procedure are displayed in Table \ref{t1} according to the guidelines for assessment of bone microstructure \cite{Bouxseinetal_JBMR_2010}. A 0.5 mm aluminium filter was used for beam filtration. Specimens were stored in phosphate buffered saline (PBS) during the scan. Reconstruction of cross sections were computed using the NRecon-Software (Bruker microCT, Kontich, Belgium). Therefore, a gaussian filter (smoothing kernel = 2) was employed. Segmentation, i.e. binarization of the data set, was performed using a locally adaptive thresholding technique (CTAn, Bruker microCT, Kontich, Belgium). 

\begin{table}
\centering
\begin{tabular}{|l|l|l|l|l|l|}
\hline
tube & tube & noise & rotation & projections & isotropic \\
voltage & current & reduction & steps ($^\circ$) & & voxel side \\
(kVp) & ($\mu$A) & (frame  & & & length ($\mu$m) \\
& & averaging) & & &  \\
\hline
45 & 175 & 4 & 0.3 & 800 & 7.1 \\
\hline
\end{tabular}
\caption{$\mu$CT parameter}
\label{t1}
\end{table}
\section{Results}
\label{sec4}

We now apply our model to a segment of a trabecular bone, which is obtained from tomography data of a sheep vertebra. Instead of the linear growth law used for validation we consider eq. (\ref{V eq2}) with a lazy zone determined by $T = 0.2$. Volumetric compression was chosen as stimulus and parameters were set to $\epsilon = 0.003$, $\gamma = 10$, $k = 1$, $d = 0.01$, $\alpha = 690$, $\beta = 1$. The elastic properties are chosen as $E = 6.829 GPa$ and $\nu = 0.33$, see \cite{Ruimermanetal_ABME_2005,Mueller2011,Mueller2014}. The cubic region $\Omega$ is defined by $L_{\rm x} = L_{\rm y} = L_{\rm z} = 2.84 mm$. To account for the dominant loading in z-direction in the animal, the applied force is set twice as high as in the other directions $F_{\rm x} = F_{\rm y} = 8 N$ and $F_{\rm z} = 16 N$. The computations are done in parallel on 24 cores with approximately 7.5 Mio degrees of freedom in each time step. The computational time for each setting was approximately 1 hour. 

\begin{figure}
$t=0$ \hspace{3.3cm} $t=0.01$ \hspace{2.6cm} $t=0.20$ \\
\includegraphics[width=0.31\textwidth,trim={145pt 30pt 165pt 120pt},clip]{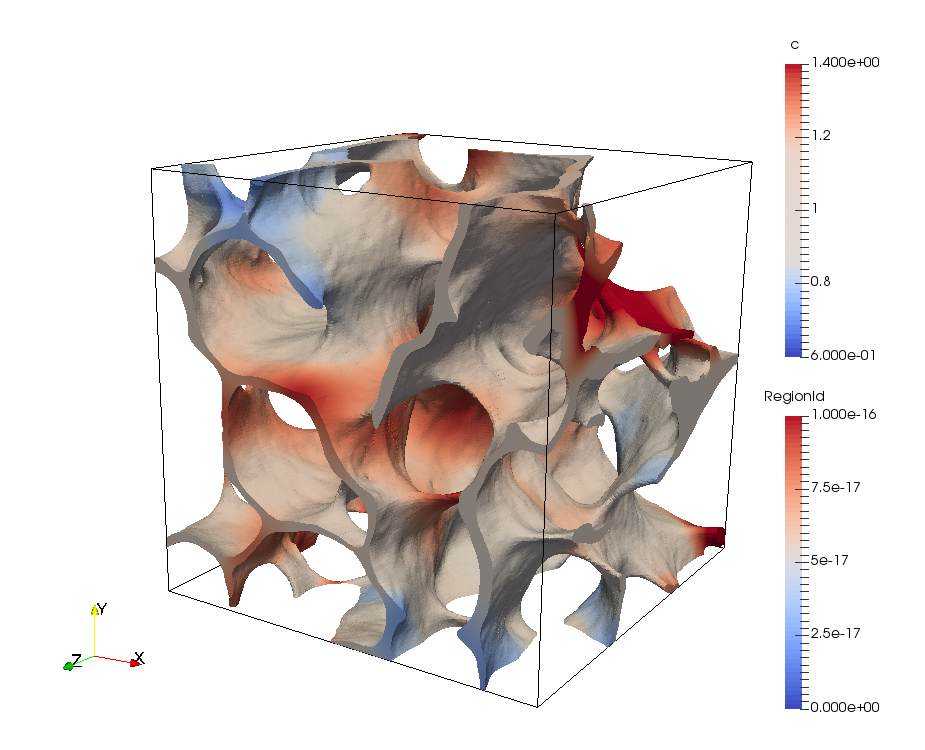} 
\includegraphics[width=0.31\textwidth,trim={145pt 30pt 165pt 120pt},clip]{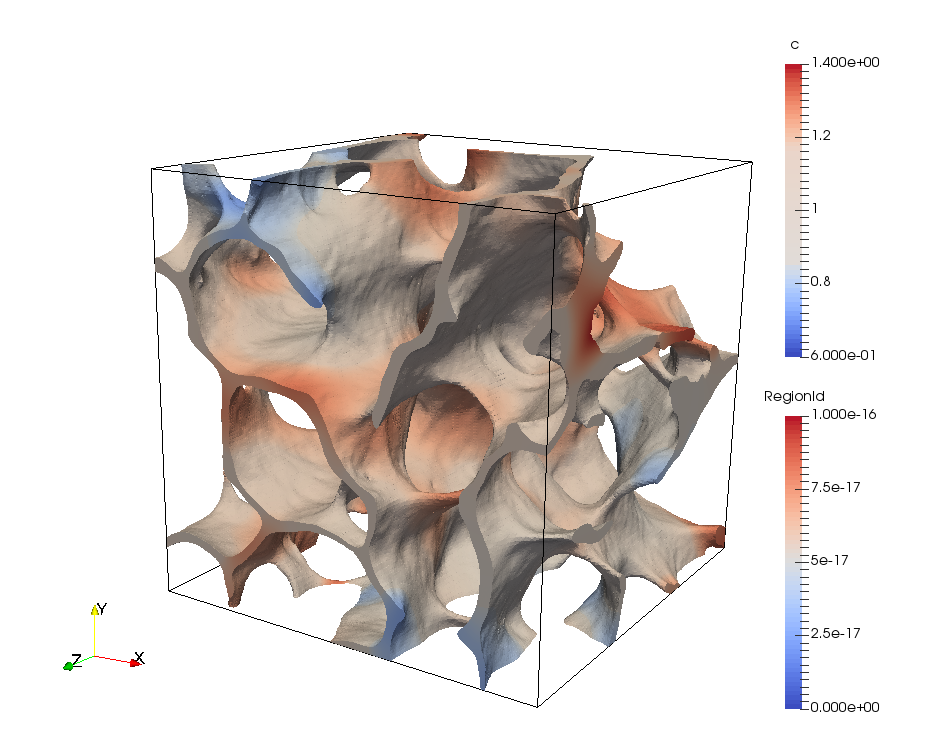} 
\includegraphics[width=0.31\textwidth,trim={145pt 30pt 165pt 120pt},clip]{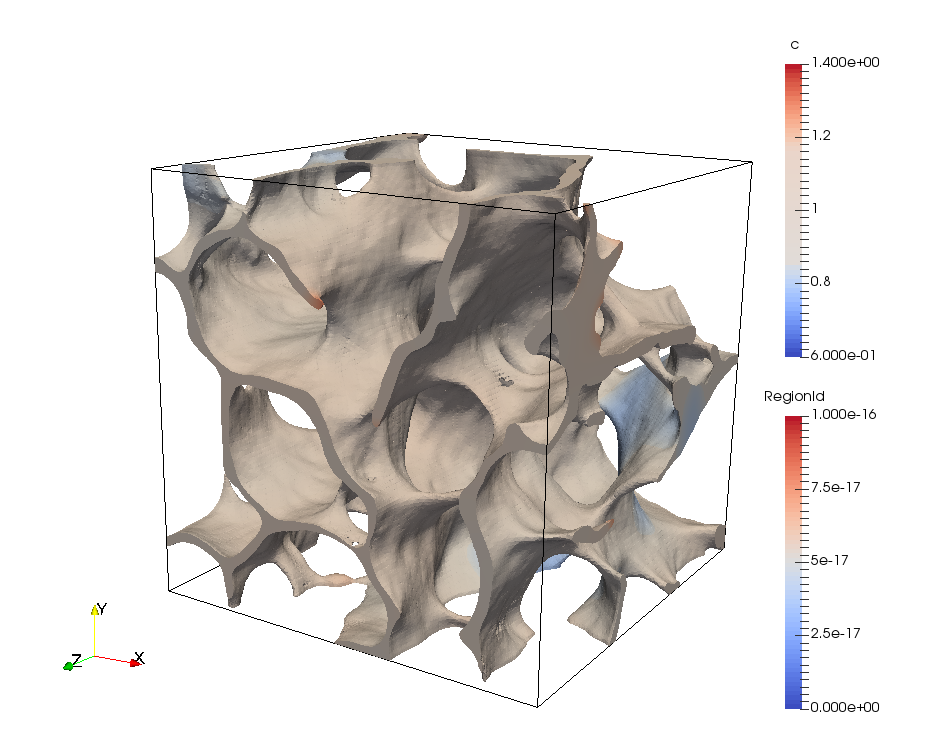} 
\includegraphics[width=0.045\textwidth]{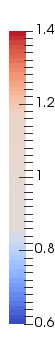} \\
\caption{Evolution of bone morphology with standard parameters. The color coding corresponds to the scaled concentration $\alpha c$.}
\label{fig:standard evolution}
\end{figure}

Fig. \ref{fig:standard evolution} shows the evolution obtained with these parameters. The evolution is shown up to a state for which the main morphological changes have been completed and the concentration $c$ has been mainly equilibrated. A change in morphology is hardly visible, however the computed average microstrain in $\Omega_{bone}$, $\sqrt{<|\epsilon_{zz}|>} = 2004$, with $\epsilon = 0.5 (\nabla \mathbf{u} + (\nabla \mathbf{u})^T)$ agrees well with physiological strains measured in bone \cite{Ehrlichetal_OI_2002}.

\begin{figure}
$F_x = 32 N$, $F_y = 8 N$, $F_z = 16 N$ \\
$t=0$ \hspace{3.3cm} $t=0.01$ \hspace{2.6cm} $t=0.20$ \\
\includegraphics[width=0.31\textwidth,trim={145pt 30pt 165pt 120pt},clip]{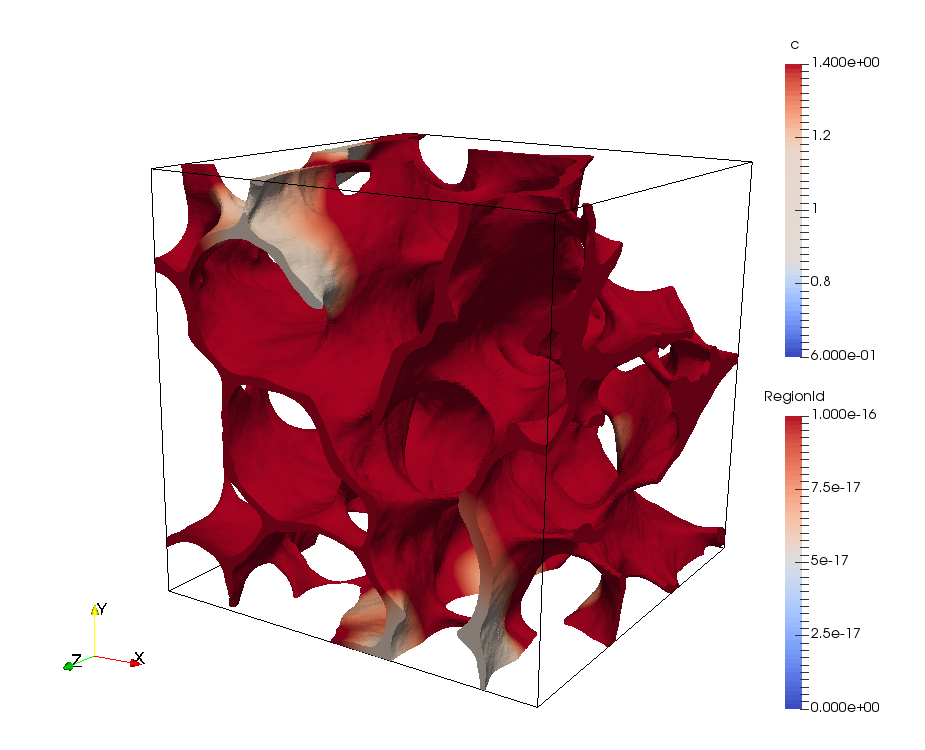} 
\includegraphics[width=0.31\textwidth,trim={145pt 30pt 165pt 120pt},clip]{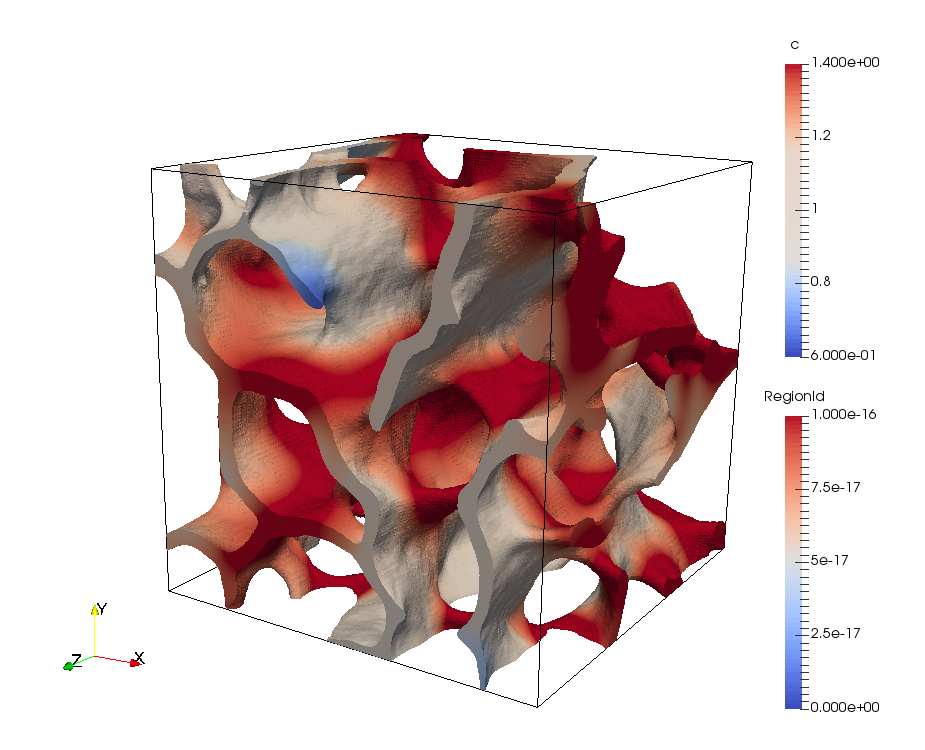} 
\includegraphics[width=0.31\textwidth,trim={145pt 30pt 165pt 120pt},clip]{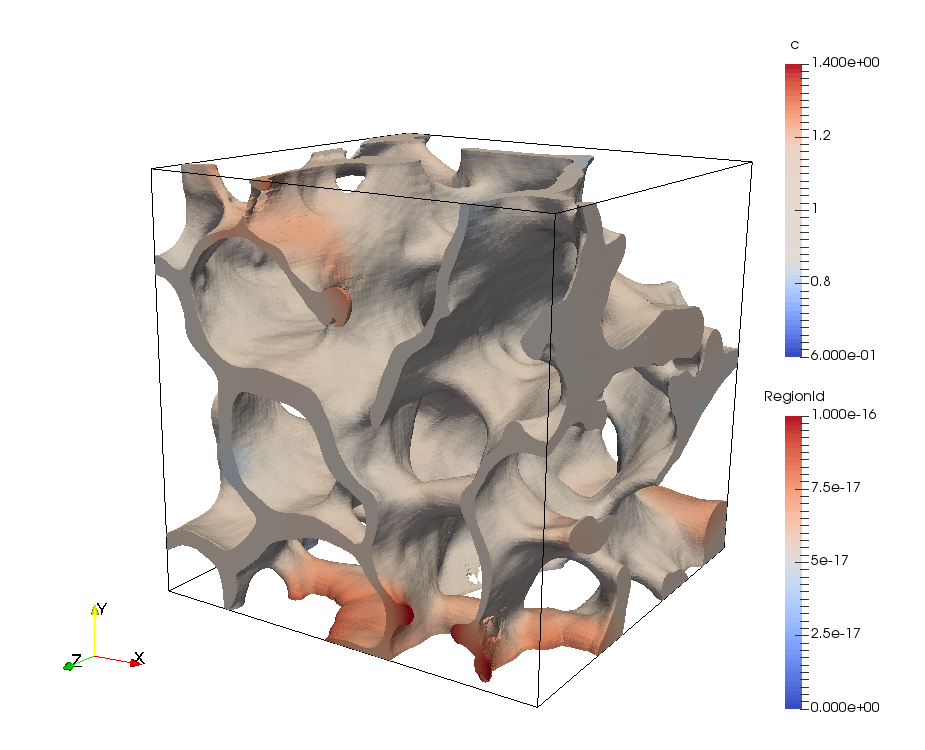} 
\includegraphics[width=0.045\textwidth]{colorbar.png} \\
$F_x = 8 N$, $F_y = 32 N$, $F_z = 16 N$ \\
$t=0$ \hspace{3.3cm} $t=0.01$ \hspace{2.6cm} $t=0.20$ \\
\includegraphics[width=0.31\textwidth,trim={145pt 30pt 165pt 120pt},clip]{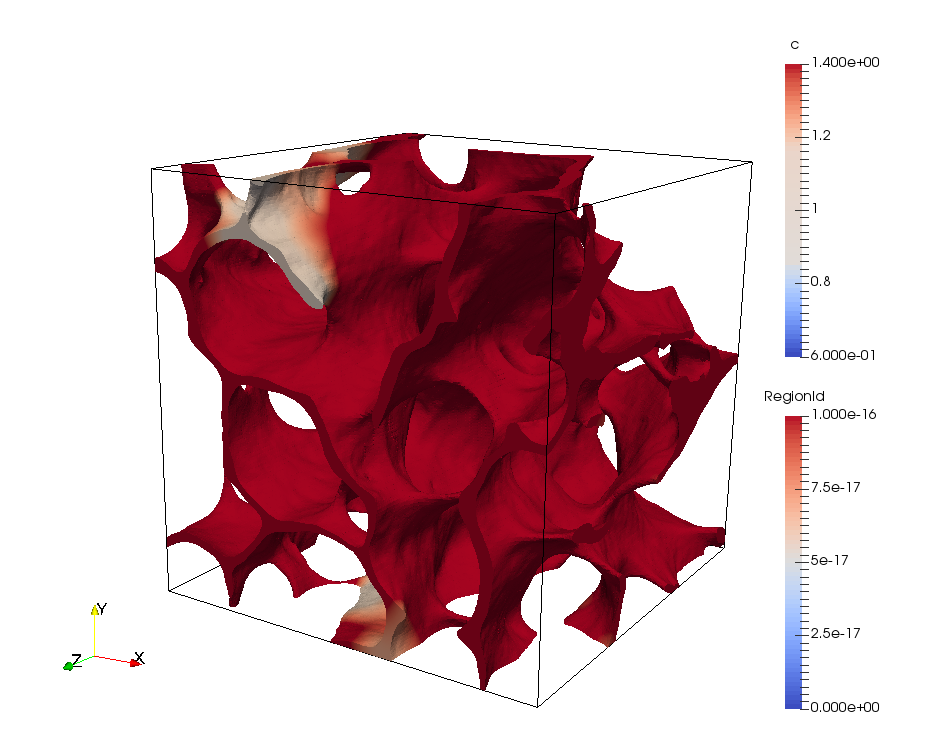} 
\includegraphics[width=0.31\textwidth,trim={145pt 30pt 165pt 120pt},clip]{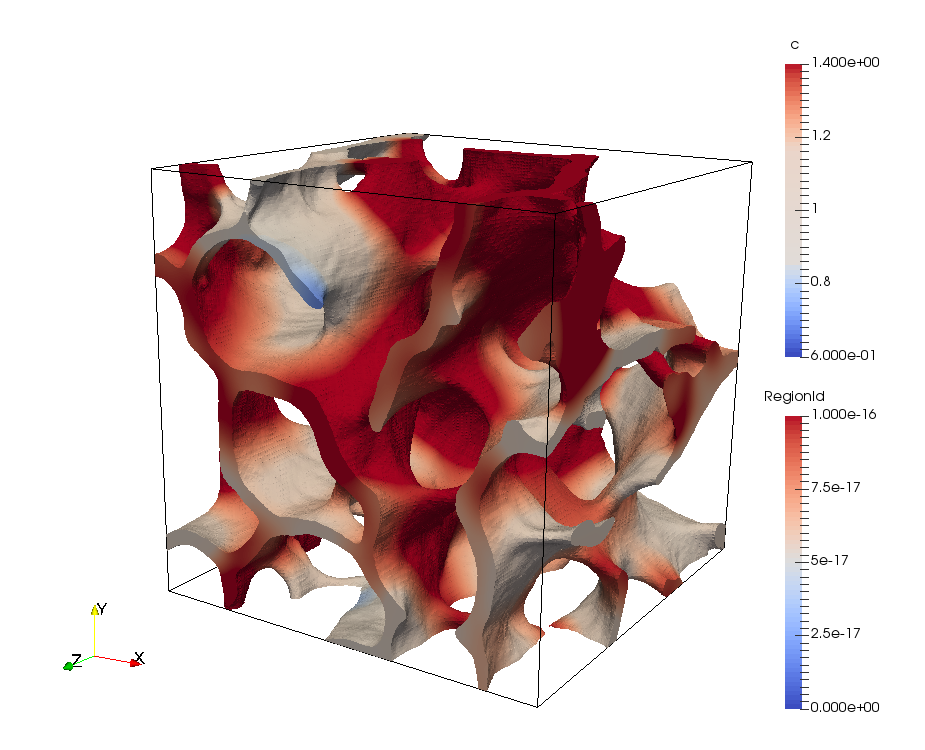} 
\includegraphics[width=0.31\textwidth,trim={145pt 30pt 165pt 120pt},clip]{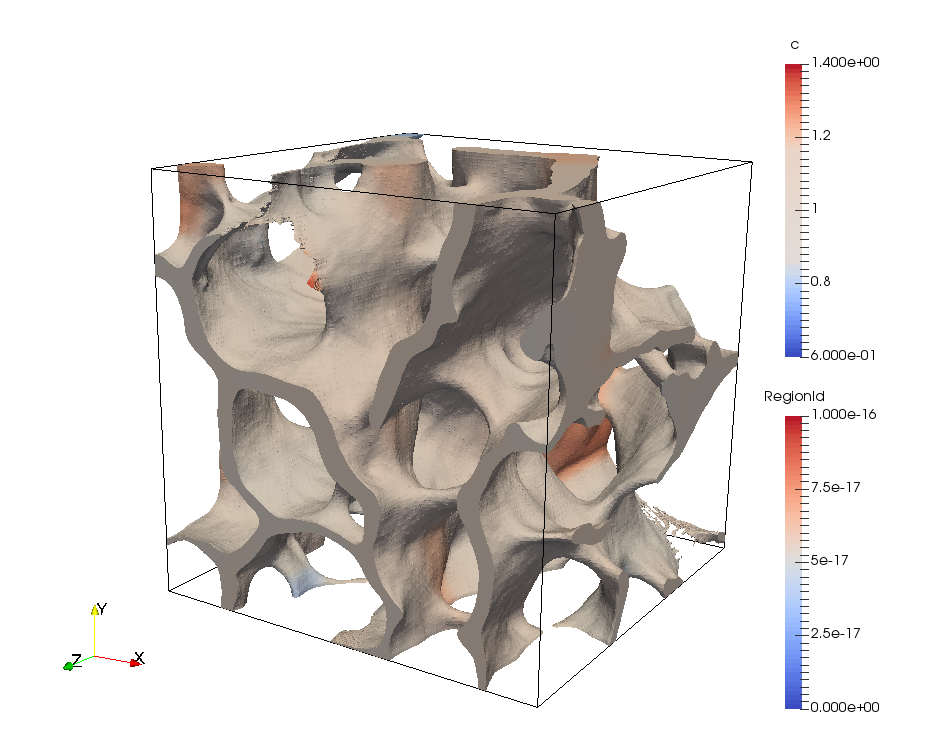} 
\includegraphics[width=0.045\textwidth]{colorbar.png} \\
$F_x = 8 N$, $F_y = 8 N$, $F_z = 64 N$ \\
$t=0$ \hspace{3.3cm} $t=0.01$ \hspace{2.6cm} $t=0.20$ \\
\includegraphics[width=0.31\textwidth,trim={145pt 30pt 165pt 120pt},clip]{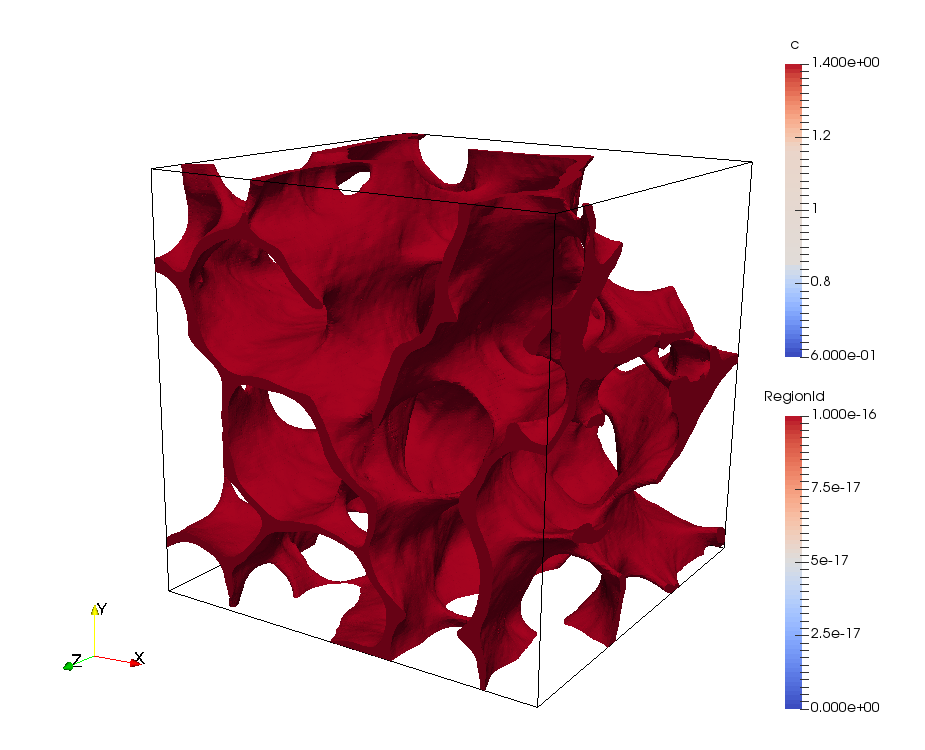} 
\includegraphics[width=0.31\textwidth,trim={145pt 30pt 165pt 120pt},clip]{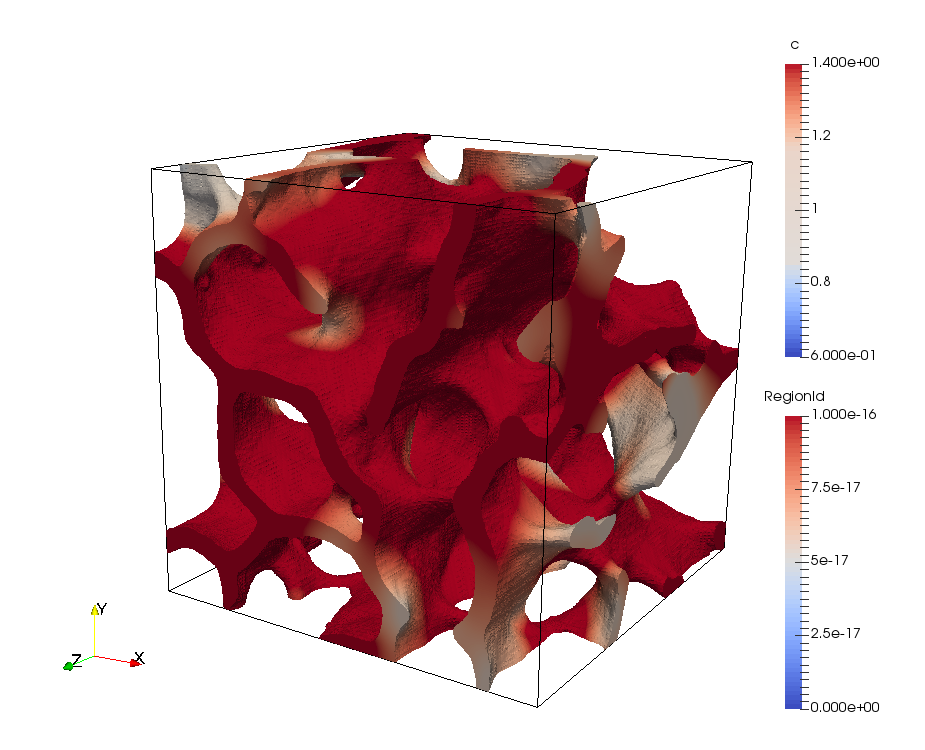}
\includegraphics[width=0.31\textwidth,trim={145pt 30pt 165pt 120pt},clip]{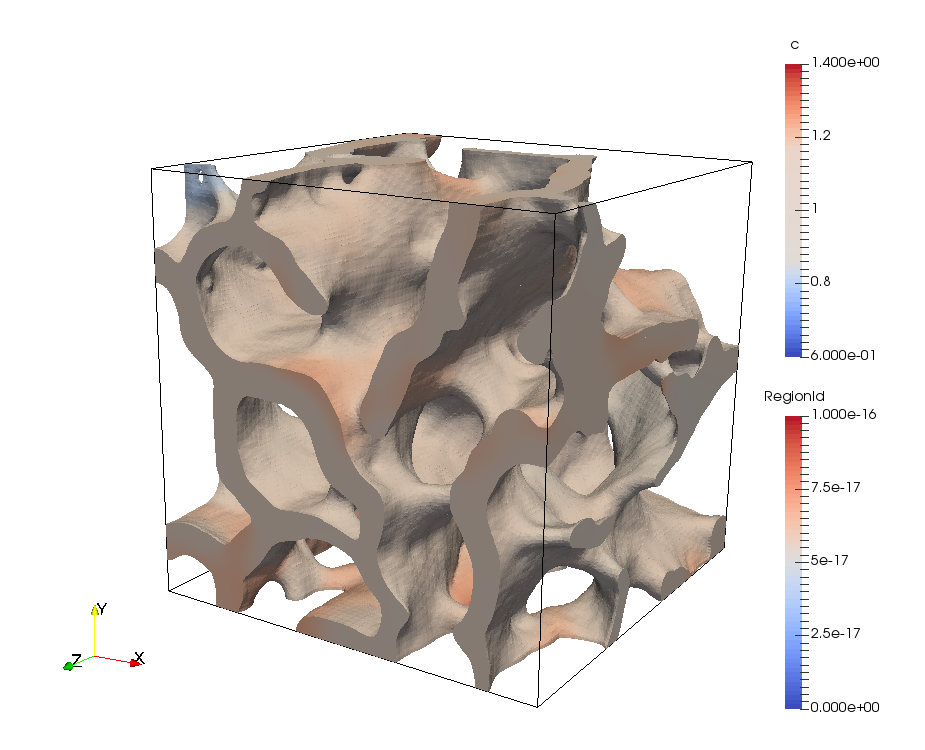} 
\includegraphics[width=0.045\textwidth]{colorbar.png} \\
\caption{Evolution of the bone morphology with four fold force in one direction. The color coding corresponds to the scaled concentration $\alpha c$.}
\label{fig:evolution 4}
\end{figure}

\begin{figure}
\center
\includegraphics[width=0.3\textwidth,trim={150pt 30pt 150pt 80pt},clip]{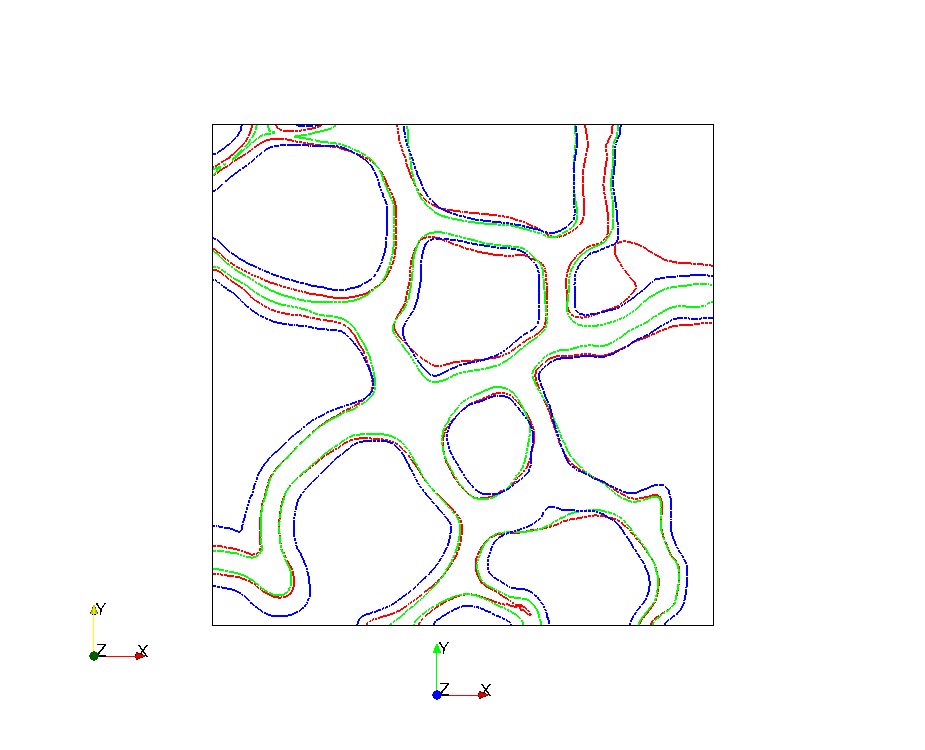}
\includegraphics[width=0.3\textwidth,trim={150pt 30pt 150pt 80pt},clip]{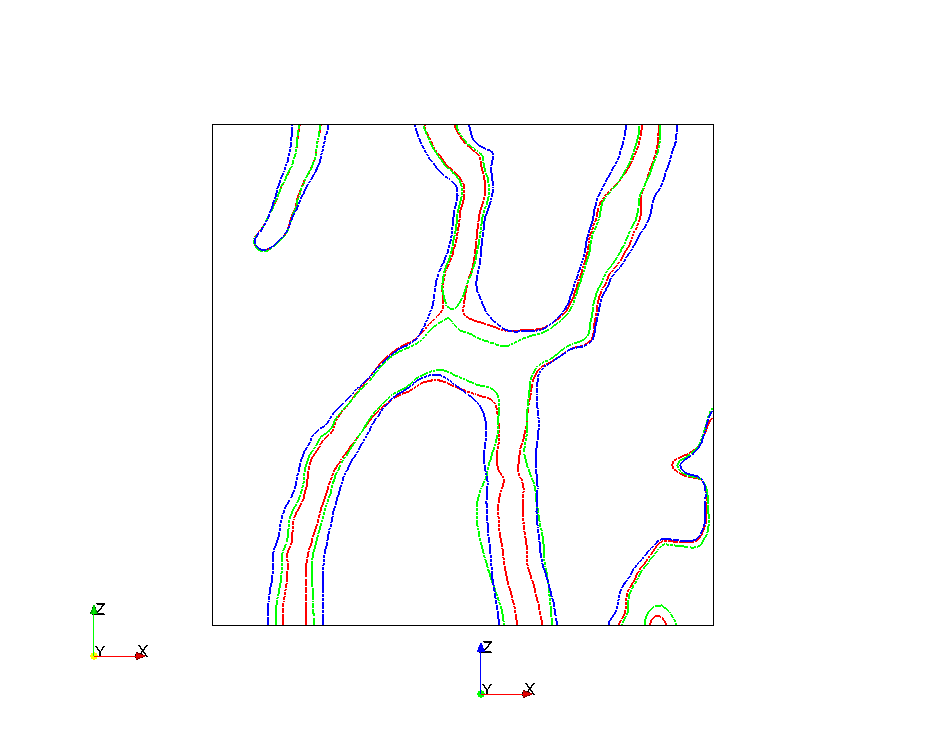}
\includegraphics[width=0.3\textwidth,trim={150pt 30pt 150pt 80pt},clip]{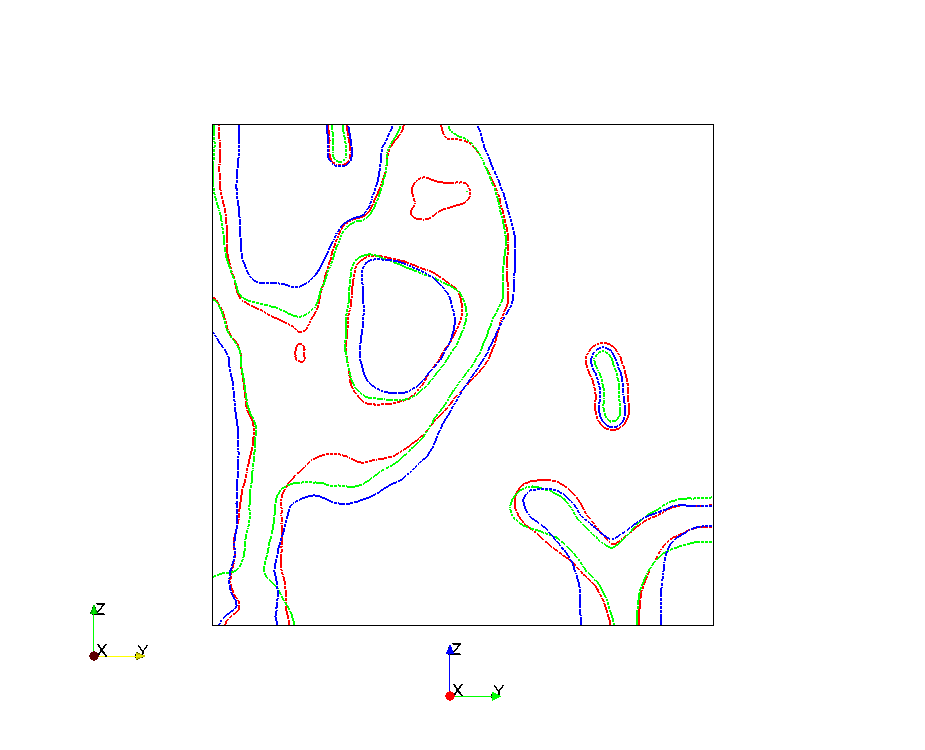}
\caption{Comparison of the bone morphology along the xy-, xz- and yz-plane (from left to right) at $t=0.20$. Applied force is multiplied by a factor of 4 in direction x (red), y (green) or z (blue).}
\label{fig:slice}
\end{figure}

To highlight the morphological adaption to anisotropic forces we consider compression forces which are increased by a factor 4 in one direction. Fig.~\ref{fig:evolution 4} shows the results. The stronger force leads to larger values for $c$ and the adaption of the morphology strongly depending on the direction of the increased force. The change in morphology is clearly visible with regions which are formed and regions which are resorbed. The dependency of the morphology on the direction of the increased force is highlighted in Fig.  \ref{fig:slice}, which shows slices of the bone morphology along the xy-, xz- and yz-plane, through the center of the domain. This visualization clearly shows that structures grow in the direction of the enhanced (4-fold) force and thus provide a proof of consistency of the modeling approach. Further validation would require in vivo $\mu$CT data of the trabecular bone segment and a calibration of the applied forces. Approaches in this direction can be found in \cite{Schulteetal_Bone_2011,Schulteetal_Bone_2013}, which could reproduce changes in bone volume fraction and other global parameters of bone structure but failed to reproduce local bone formation and resorption. One reason for this discrepancy, which is reported in \cite{Schulteetal_Bone_2011,Schulteetal_Bone_2013} are local areas of strong thickening and bone resorption in the experimental images in contrast to more homogeneous layers in their simulations. Fig.~\ref{fig:slice} clearly shows non-homogeneous morphology changes, see e.g. the level curves for an increased force in x-direction (red curves) in the first and third figure. However, due to lack of in vivo data for the considered bone segment, further validation has to be postponed and we here can only conclude the qualitative consistence of our approach.

\section{Discussion}
\label{sec6}

We have introduced a continuous modeling approach for bone remodeling which has the potential to combine different scales in an efficient way. In the current form it combines elastic responds of the trabecular bone structure to an applied force, the concentration of signaling molecules within the bone and a mechanism how this concentration at the bone surface is used for local bone formation and resorption. In contrast to previous modeling approaches, using micro finite element analysis, the bone morphology is implicitly described by a time-dependent phase field function. This not only leads to a more accurate model, as the artificial voxel-roughness of the bone surface can be avoided, but also to a drastic reduction of system size and required computing time. The goal of this paper is to provide a minimal model for bone remodeling which demonstrates the advantages of the phase field description. We therefore not only introduce the model, but also quantitatively compare the results with established micro finite element approaches on simple geometries and consider the bone morphology within a segment of $2.84mm^3$ obtained from $\mu$CT data of a sheep vertebra with realistic parameters. Systematic studies with an enhanced force in one direction clearly demonstrate that the structures grow in the direction of the enhanced force and lead to strong local variations in thickness. These results clearly demonstrate the applicability of the phase field approach. However, quantitative validation would require in vivo $\mu$CT data of the trabecular bone segment and a calibration of the applied forces, which are currently not available. But already without such a validation the approach can be used to provide a deeper understanding of the mechanisms underlying bone remodeling. A possible extension considers the incorporation of
osteoblast and osteoclast concentrations on the bone surface and their biochemical signaling, which will allow to compute the influence of various signaling molecules on the bone morphology. The ability to predict changes in bone morphology might eventually lead to a better prediction of individual fracture risk in osteoporotic patients or to improved implant materials. The introduced phase field description is ideally suited for this task, as it drastically reduces the computational cost, allows for extensions of additional effects and does only require standard numerical methods, which can be parallelized to deal with significantly larger systems.  

\vspace*{0.5cm}
\noindent {\bf Acknowledgement:} The authors acknowledge support by the German Research Foundation (DFG) through SFB/TRR 79 "Materials for tissue regeneration within systemically altered bone" (project M8, Z3, T1). We used computing resources provided by JSC within project HDR06 as well as resources provided by ZIH at TU Dresden.

\bibliography{biball}

\end{document}